\documentclass[aps,pra,reprint,superscriptaddress]{revtex4-2}
\usepackage{amsmath}
\usepackage{amssymb}
\usepackage{mathtools}
\usepackage{xcolor}
\usepackage{graphicx}
\usepackage[normalem]{ulem}

\newcommand{\ket}[1]{\left\vert#1\right\rangle}
\newcommand{\bra}[1]{\left\langle#1\right\vert}
\newcommand{\inner}[2]{\left\langle#1\vert#2\right\rangle}

\DeclareMathOperator{\Ai}{Ai}

\begin{document}

\title{Wave-function engineering via conditional quantum teleportation with non-Gaussian entanglement resource}


\author{Warit Asavanant}
\email{warit@alice.t.u-tokyo.ac.jp}
\author{Kan Takase}
\author{Kosuke Fukui}
\author{Mamoru Endo}
\author{Jun-ichi Yoshikawa}
\author{Akira Furusawa}
\email{akiraf@ap.t.u-tokyo.ac.jp}
\affiliation{Department of Applied Physics, School of Engineering, The University of Tokyo, 7-3-1 Hongo, Bunkyo-ku, Tokyo 113-8656, Japan}


\date{\today}

\begin{abstract}
We propose and analyze a setup to tailor the wave functions of the quantum states. Our setup is based on the quantum teleportation circuit, but instead of the usual two-mode squeezed state, two-mode non-Gaussian entangled state is used. Using this setup, we can generate various classes of quantum states such as Schr{\"o}dinger cat states, four-component cat states, superpositions of Fock states, and cubic phase states. These results demonstrate the versatility of our system as a state generator and suggest that conditioning using homodyne measurements is an important tool in the generations of the non-Gaussian states in complementary to the photon number detection.
\end{abstract}


\maketitle

\section{Introduction}
Quantum information processing is widely researched with expectations of broad applications \cite{nielsen00}. Among many candidates, continuous-variable (CV) optical systems are currently one of the most promising platforms in terms of scalability. In CV optical systems, resources for measurement-based quantum computation (MBQC) \cite{PhysRevLett.86.5188,PhysRevLett.97.110501}---the cluster states---have been generated in a scalable fashion using time-domain multiplexing method \cite{Yokoyama2013,doi:10.1063/1.4962732,Asavanant373,Larsen369} and frequency-domain multiplexing method \cite{PhysRevLett.112.120505}, and basic operations on time-domain-multiplexed cluster states have been recently demonstrated \cite{2020arXiv200611537A,2020arXiv201014422L}. To achieve universal MBQC, however, non-Gaussian elements have to be added to the cluster states \cite{PhysRevLett.82.1784,PhysRevLett.88.097904}.

A direct way to add the non-Gaussian elements is adding non-Guassian measurements to the cluster state. This, however, is experimentally difficult, especially in optical systems, where direct non-Gaussian measurements are usually either probabilistic or require the non-available strong nonlinearity. A more viable option is to use ancillary non-Gaussian states (such as cubic phase states) and realize non-Gaussian operations via gate teleportation protocol \cite{Gottesman1999,PhysRevA.64.012310}. In addition, non-Gaussian states also have many other applications. For example, a superposition of coherent states---commonly known as a Schr{\"o}dinger cat state---have applications in quantum computation \cite{PhysRevA.59.2631,PhysRevA.68.042319,PhysRevLett.100.030503,PhysRevLett.105.053602}, quantum communication \cite{PhysRevA.65.042320,Sangouard:10}, and quantum error correction \cite{Vasconcelos:10,Hastrup:20,PhysRevA.97.022341}. Another non-Gaussian state called Gottesman-Kitaev-Preskill (GKP) qubit is currently the most promising logical qubit for fault-tolerant CV quantum computation \cite{PhysRevA.64.012310,PhysRevLett.112.120504,PhysRevX.8.021054,PhysRevA.100.010301,PhysRevLett.123.200502,PhysRevResearch.2.023270,PhysRevA.101.032315}.

Despite their varieties, optical generations of non-Gaussian states are mostly based on the same idea: the quantum state is first represented in the photon number basis (Fock basis), i.e.\ $\ket{\psi}=\sum_{n=0}^{\infty}c_{n}\ket{n}$, and the target state is obtained by truncating the superposition below a certain maximum photon number, $\ket{\psi_\textrm{target}}\approx\sum_{n=0}^{n_{\textrm{max}}}c_{n}\ket{n}$. Then, we can find a setup that consists of squeezed light sources, linear optics, and photon number detectors that herald $\ket{\psi_\textrm{target}}$ when particular results are detected by the photon number detectors. Based on this idea, generations of various non-Gaussian states have been explored \cite{PhysRevA.101.032315,PhysRevLett.88.250401,PhysRevLett.88.113601,PhysRevLett.123.113603,Bimbard2010,PhysRevA.86.043820,Yukawa:13,PhysRevA.55.3184,PhysRevLett.115.023602,Hastrup:20,PhysRevA.82.031802,Ourjoumtsev83,PhysRevLett.97.083604,PhysRevLett.105.053602,Wakui:07}. In this approach, however, the generation system is highly dependent on the target states and the generation systems tend to become more complex when the maximum photon number increases. Also, although these generations are formulated based on Fock basis, for CV quantum states, phase space representation in quadrature basis is also a natural representation. As the examples of previous researches in this direction, generation of non-Gaussian state by the implementation of the quadrature operator \cite{PhysRevA.90.013804}, generation and amplification of Schr{\"o}dinger cat states using conditional homodyne measurement \cite{Ourjoumtsev2007,Sychev2017}, and high-rate Schr{\"o}dinger cat state generation formulated using the wave function picture \cite{PhysRevA.103.013710} have been studied. Even so, the non-Gaussian state generation using the quadrature basis has remained largely unexplored.

In this paper, we present a methodology to tailor the wave functions of the quantum states using quantum teleportation circuit and non-Gaussian entanglement resource. The idea of our protocol is based on gate teleportation protocol \cite{Gottesman1999,PhysRevA.64.012310}, where we have designed the non-Gaussian two-mode resource states to be equivalent to EPR state---a resource state for CV quantum teleportation---with quadrature operators (such as $\hat{x}$) acting on one of its mode in the ideal limit. As the quadrature operators and their polynomials are non-unitary, we need to use the conditional quantum teleportation instead of the conventional unconditional quantum teleportation in the state preparation using our protocol. In this case, the measurement results of the Bell measurements herald the generated states. In addition, this teleportation-based architecture possesses high affinity with the time-domain-multiplexing method \cite{PhysRevA.83.062314}, making it possible to realize our scheme in a scalable fashion. Our results demonstrate a programmable and scalable non-Gaussian quantum state generator that utilizes the quadrature basis.

The paper is structured as follows. In Sec.\ \ref{sec:notation}, we define the basic notations. In Sec.\ \ref{sec:methodology}, we analyze our methodology. In Sec.\  \ref{sec:simulation}, we show the actual generations of various non-Gaussian states: Schr{\"o}dinger cat states, four-component cat states, superpositions of Fock states, and cubic phase states. We discuss success rate and realistic implementation of our architecture in Sec.\ \ref{sec:discussion}. Finally, we conclude our paper in Sec.\ \ref{sec:conclusion}.

\section{Notations\label{sec:notation}}

In CV systems, the quadrature operators $\hat{x}$ and $\hat{p}$ satisfy commutation relation: $[\hat{x},\hat{p}]=i$, where we use $\hbar=1$. The quadrature operators are related to the annihilation and creation operators via
\begin{align}
\hat{x}&=\frac{1}{\sqrt{2}}(\hat{a}+\hat{a}^\dagger),\label{eq:x_def}\\
\hat{p}&=\frac{1}{i\sqrt{2}}(\hat{a}-\hat{a}^\dagger),\label{eq:p_def}
\end{align}
with $[\hat{a},\hat{a}^\dagger]=1$, and the effect of the annihilation and the creation operators on the Fock basis $\ket{n}$ are
\begin{align}
\hat{a}\ket{n}&=\sqrt{n}\ket{n-1},\\
\hat{a}^\dagger\ket{n}&=\sqrt{n+1}\ket{n+1}.
\end{align} 

For a quantum states $\ket{\psi}$, the wave function in $x$ and $p$ are given by
\begin{align}
\psi(x)&=\inner{x}{\psi},\\
\tilde{\psi}(p)&=\inner{p}{\psi},
\end{align}
respectively. When quadrature operators act on the quantum states, they transform the wave functions into a new (unnormalized) wave function given by
\begin{align}
\langle x\vert\hat{x}\ket{\psi}&=x\psi(x),\\
 \langle x\vert\hat{p}\ket{\psi}&=-i\frac{\textrm{d}\,}{\textrm{d}x}\psi(x),\\
\langle p\vert\hat{x}\ket{\psi}&=i\frac{\textrm{d}\,}{\textrm{d}p}\tilde{\psi}(p),\\
\langle p\vert\hat{p}\ket{\psi}&=p\tilde{\psi}(p).
\end{align}

In the quantum teleportation circuit, the ideal two-mode resources are called EPR states and their unnormalized form are given by
\begin{align}
\ket{\textrm{EPR}}&=\int\textrm{d}x\,\ket{x}_{1}\ket{x}_{2}=\sum_{n=0}^{\infty}\ket{n}_{1}\ket{n}_{2}.
\end{align}
Note that unless stated otherwise, we assume that the ranges of the integrations are always from $-\infty$ to $\infty$. In the physical setting, two-mode squeezed state (TMSS) is used instead of the EPR state and it can be represented as
\begin{align}
\begin{split}
\ket{\textrm{TMSS}}&=\int\textrm{d}x_{1}\textrm{d}x_{2}\,\Psi_\textrm{TMSS}(x_{1},x_{2})\ket{x}_{1}\ket{x}_{2}\\
&=\sqrt{1-\eta^{2}}\sum_{n=0}^{\infty}\eta^{n}\ket{n}_{1}\ket{n}_{2},
\end{split}
\end{align}
with
\begin{align}
\begin{split}
\Psi_\textrm{TMSS}(x_{1},x_{2})=&\exp\left[-\frac{e^{2r}}{2}\left(\frac{x_{1}-x_{2}}{\sqrt{2}}\right)^2\right]\\
&\times\exp\left[-\frac{e^{-2r}}{2}\left(\frac{x_{1}+x_{2}}{\sqrt{2}}\right)^2\right]\label{eq:TMSSwave}
\end{split}\\
\eta=&\tanh r,
\end{align}
where $r$ is the squeezing parameter of the initial squeezed vacua used in the generation of TMSS and is assumed to be initially equal for both modes. We can see that in the limit of the infinite squeezing, the first term approaches $\delta(x_{1}-x_{2})$, while the second term is a Gaussian envelope that becomes broader as $r$ increases.

TMSS can be generated by mixing two orthogonal squeezed lights on a beam splitter:
\begin{align}
\ket{\textrm{TMSS}}=\hat{B}\hat{S}_{1}(-r)\hat{S}_{2}(r)\ket{0}_1\ket{0}_2.
\end{align}
$S(r)\coloneqq\exp\left[\frac{r}{2}(\hat{a}^2-\hat{a}^{\dagger 2})\right]$ is a squeezing operator where $r>0$ ($r<0$) corresponds to squeezing in $\hat{x}$ ($\hat{p}$) quadrature. $\hat{B}$ is an operator of 50:50 beamsplitter interaction which transforms the annihilation operator of the two modes as
\begin{align}
\hat{B}^\dagger\begin{pmatrix}
\hat{a}_1\\
\hat{a}_2
\end{pmatrix}\hat{B}=\frac{1}{\sqrt{2}}\begin{pmatrix}
1&-1\\
1&1
\end{pmatrix}\begin{pmatrix}
\hat{a}_1\\
\hat{a}_2
\end{pmatrix}.
\end{align}

\section{Proposed setup\label{sec:methodology}}
\begin{figure}
\includegraphics[width=\columnwidth]{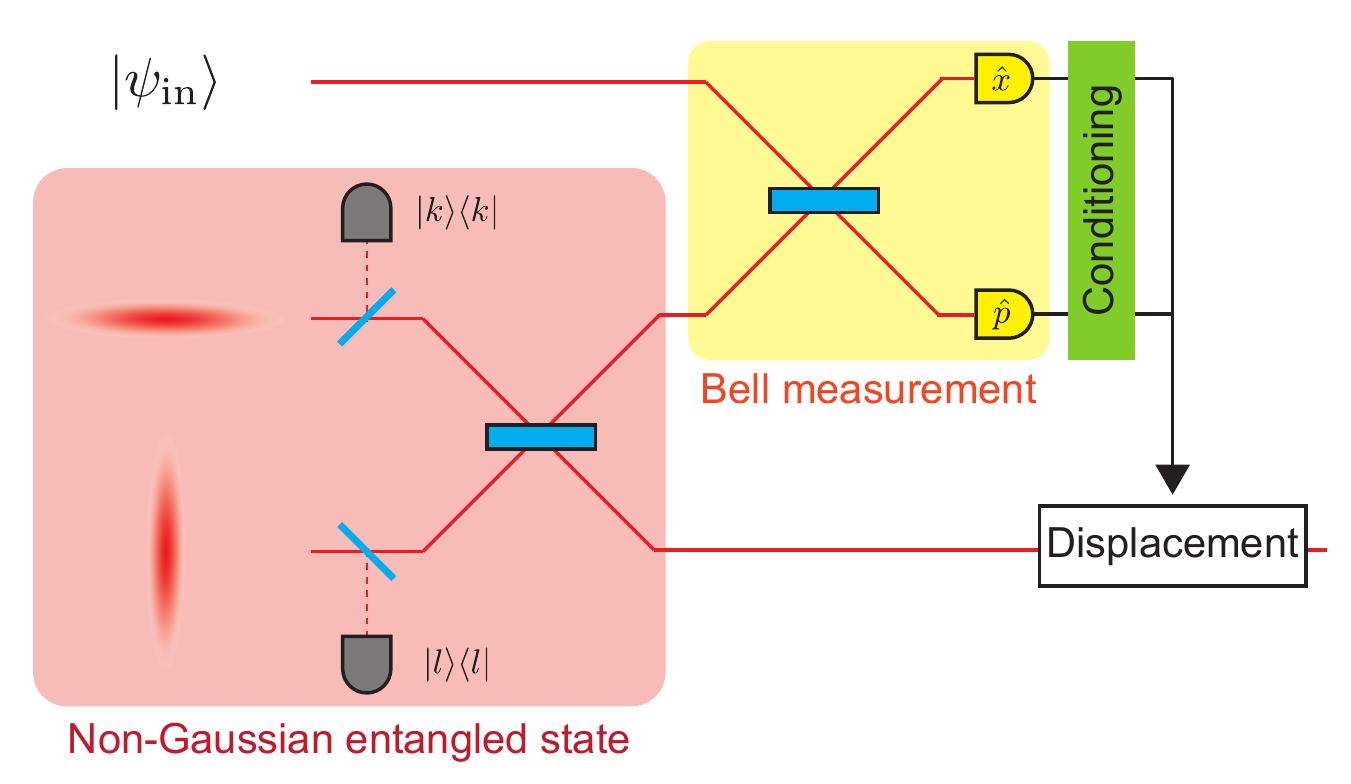}%
\caption{Schematic diagram of the proposed setup. Our setup is a quantum teleportation circuit where the usual two-mode squeezed state (TMSS) are replaced with the states which we will call non-Gaussian entangled state (NGES).  One of the possible generation methods used in the analysis in this paper is that by photon subtractions on the squeezed states used in the generation of the TMSS. The success of the protocol is conditioned by certain measurement results at the Bell measurements.\label{fig:setup}}
\end{figure}

In this section, we analyze the proposed setup. The generalized form of our setup is shown in Fig.\ \ref{fig:setup}. This setup consists of photon-subtracted two-mode entanglement and conditioning quantum teleportation with that resource. Let us look at each component.

\subsection{Photon-subtracted TMSS}
We consider what happens when photon subtractions are combined with TMSS. The main reason we consider this type of state is that the quadrature operators can be written as superpositions between the annihilation and creation operators. Therefore, as we will shortly show, by implementing the photon subtraction before the 50:50 beamsplitter, we can easily implement the superposition of the annihilation and the creation operator, i.e. quadrature operators, on one of the modes of the TMSS. First, let us consider subtraction of $k$ photons from one of the mode of the TMSS. This state can be written as
\begin{align}
(\hat{a}_{1})^k\ket{\textrm{TMSS}}=\sum_{n=k}^\infty\eta^{n}\sqrt{\frac{n!}{(n-k)!}}\ket{n-k}_{1}\ket{n}_{2}.
\end{align}
Note that we will omit the normalization factor for the simplicity. The above equation can be equivalently written as
\begin{align}
\begin{split}
(\hat{a}_{1})^k\ket{\textrm{TMSS}}&=(\hat{a}_{2}^\dagger)^{k}\sum_{n=k}^\infty\eta^{n}\ket{n-k}_{1}\ket{n-k}_{2}\\
&=(\eta\hat{a}^{\dagger}_{2})^k\ket{\textrm{TMSS}}.\label{eq:subtractTMSS}
\end{split}
\end{align}
This means that photon subtraction in a single mode can be considered as photon addition in another mode with additional factor $\eta$.

Now let us return to the non-Gaussian entanglement in Fig.\ \ref{fig:setup}. The non-Gaussian entangled state $\ket{\textrm{NGES}(k,l)}$ is given by
\begin{align}
\ket{\textrm{NGES}(k,l)}=\hat{B}(\hat{a}_{1})^k(\hat{a}_{2})^l\hat{S}_1(-r)\hat{S}_2(r)\ket{0}_1\ket{0}_2.
\end{align}
Using the beamsplitter transformation, this state can be transformed into
\begin{align}
\ket{\textrm{NGES}(k,l)}=\left(\frac{\hat{a}_{1}+\hat{a}_{2}}{\sqrt{2}}\right)^{k}\left(\frac{\hat{a}_{1}-\hat{a}_{2}}{\sqrt{2}}\right)^{l}\ket{\textrm{TMSS}}.
\end{align}
Using Eq.\ \eqref{eq:subtractTMSS}, we get
\begin{align}
\ket{\textrm{NGES}(k,l)}=\hat{f}_{k,l}(\eta)\ket{\textrm{TMSS}},\label{eq:NGES}
\end{align}
where 
\begin{align}
\hat{f}_{k,l}(\eta)\equiv\frac{1}{\sqrt{2^{k+l}}}\sum_{j=0}^{k+l}c_{j}^{k,l}\hat{a}_{2}^{k+l-j}(\eta\hat{a}_{2}^\dagger)^j\label{eq:fkl}
\end{align}
and $c_{j}^{k,l}$ is a coefficient of the polynomial
\begin{align}
(a+b)^k(a-b)^l=\sum_{j=0}^{k+l}c_{j}^{k,l}a^{j}b^{k+l-j}.
\end{align}

Therefore, by implementing photon subtractions on the initial squeezed states, we can implement a polynomial of annihilation and creation operators on one of the modes of the TMSS. This is because the subtraction before the beamsplitter is equivalent to the superposition between the photon subtraction on each mode of the TMSS, which is also equivalent superposition of the photon subtraction and addition on one of the modes. In that sense, the physical intuition of our method is that quantum teleportation using this non-Gaussian entanglement is roughly the same as the implementation of the polynomial $f_{kl}(\hat{a}_{2},\eta\hat{a}_{2}^\dagger)$ on the input state. The implementation of the coherent superposition of the annihilation and the creation operators on a quantum state has been studied in a different context \cite{PhysRevA.82.053812}. There are also effects from the measurement results of the bell measurements which will be discussed in Sec.\ \ref{subsec:output}.

Although Eq.\ \eqref{eq:NGES} and Eq.\ \eqref{eq:fkl} give complete characterization of NGES, it will be more convenient to define $\hat{f}_{k,l}(\eta)$ via a recursive formula. From Eq.\ \eqref{eq:NGES}, we can write down the following recursive formula
\begin{align}
\hat{f}_{k,l}&=\frac{1}{\sqrt{2}}\hat{a}\hat{f}_{k-1,l}+\frac{1}{\sqrt{2}}\eta\hat{f}_{k-1,l}\hat{a}^\dagger,\\
\hat{f}_{k,l}&=-\frac{1}{\sqrt{2}}\hat{a}\hat{f}_{k,l-1}+\frac{1}{\sqrt{2}}\eta\hat{f}_{k,l-1}\hat{a}^\dagger,
\end{align}
with $\hat{f}_{0,0}=\hat{I}$ and we dropped the index ``2'' and the dependence on $\eta$ of $\hat{f}_{k,l}(\eta)$. This can be further written using quadrature operators as
\begin{align}
\begin{split}
\hat{f}_{k,l}=&\frac{1}{2}\left(\hat{x}\hat{f}_{k-1,l}+\eta\hat{f}_{k-1,l}\hat{x}\right)\\
&+\frac{i}{2}\left(\hat{p}\hat{f}_{k-1,l}-\eta\hat{f}_{k-1,l}\hat{p}\right),
\end{split}\\
\begin{split}
\hat{f}_{k,l}=&-\frac{1}{2}\left(\hat{x}\hat{f}_{k,l-1}-\eta\hat{f}_{k,l-1}\hat{x}\right)\\
&-\frac{i}{2}\left(\hat{p}\hat{f}_{k,l-1}+\eta\hat{f}_{k,l-1}\hat{p}\right).
\end{split}
\end{align}

In the limit of $\eta\to1$, we can show that $\hat{f}_{k,0}$ is the polynomial of solely $\hat{x}$ and $\hat{f}_{0,l}$ is the polynomial of solely $\hat{p}$. This can be done as follows. Let us put $\hat{g}_{k,l}\equiv\lim_{\eta\to1}\hat{f}_{k,l}$. Then,
\begin{align}
\hat{g}_{k,l}&=\frac{1}{2}\{\hat{x},\hat{g}_{k-1,l}\}+\frac{i}{2}[\hat{p},\hat{g}_{k-1,l}],\\
\hat{g}_{k,l}&=-\frac{1}{2}[\hat{x},\hat{g}_{k,l-1}]-\frac{i}{2}\{\hat{p},\hat{g}_{k,l-1}\}.
\end{align}
As we can easily show that $\hat{g}_{1,0}=\hat{x}$ and $\hat{g}_{0,1}=-i\hat{p}$, using the recursive formula of the Hermite polynomial, we can write down
\begin{align}
\hat{g}_{k,0}&=\frac{1}{(2i)^k}H_{k}(i\hat{x}),\\ 
\hat{g}_{0,l}&=\frac{1}{(-2)^l}H_{l}(i\hat{p}),
\end{align}
where $H_{n}(\cdot)$ is the Hermite polynomials of the order $n$. The reason that both $\hat{g}_{k,0}$ and $\hat{g}_{0,l}$ are polynomial in $\hat{x}$ or $\hat{p}$ with real coefficients is because we are utilizing the phase information and the splitting ratio of the beamsplitter, whereas general arbitrary superposition of the annihilation and creation operators do not necessarily result in the real-coefficient polynomial in the quadrature operators.

As an example of $\hat{f}_{k,l}$, we list a few of them here. Note that $\hat{f}_{k,l}$ are always at most the polynomial of the order $k+l$ in $\hat{x}$ and $\hat{p}$.
\begin{align}
\hat{f}_{1,0}&=\frac{1+\eta}{2}\hat{x}+i\frac{1-\eta}{2}\hat{p}\\
\hat{f}_{0,1}&=-\frac{1-\eta}{2}\hat{x}-i\frac{1+\eta}{2}\hat{p}\\
\hat{f}_{1,1}&=-\frac{1-\eta^2}{4}(\hat{x}^2-\hat{p}^2)-i\frac{1+\eta^2}{4}(\hat{x}\hat{p}+\hat{p}\hat{x})\\
\begin{split}
\hat{f}_{2,0}&=\left[\frac{1+\eta}{2}\right]^2\hat{x}^2+\left[i\frac{1-\eta}{2}\right]^2\hat{p}^2\\
&+\frac{1}{4}(\eta^2-1+2\eta)
\end{split}
\end{align}

\subsection{Output states\label{subsec:output}}
Now that we have established that our two-mode resource is equivalent to coherent superposition of photon subtraction and photon addition acting on one of the modes of the TMSS, we look at the conditional quantum teleportation part. When we implement Bell measurement, we are projecting the mode ``in'' and ``1'' onto the displaced EPR states via the projection operator:
\begin{align}
\begin{split}
&\hat{\Pi}(m_x,m_p)=\\
&\hat{D}_{x,\textrm{in}}(m_x)\hat{D}_{p,1}(m_p)\ket{\textrm{EPR}}_{\textrm{in},1}\bra{\textrm{EPR}}\hat{D}^\dagger_{x,\textrm{in}}(m_x)\hat{D}^\dagger_{p,1}(m_p),
\end{split}\label{eq:EPRprojector}
\end{align}
where the displacement operators $\hat{D}_x(m_x)$ and $\hat{D}_p(m_p)$ transform quadrature operators as $\hat{D}_x^\dagger(m_x)\hat{x}\hat{D}_x(m_x)=\hat{x}+m_x$ and $\hat{D}_p^\dagger(m_p)\hat{p}\hat{D}_p(m_p)=\hat{p}+m_p$. Note that $m_x$ and $m_p$ are not directly the measurement results of the two homodyne measurements but are related to that by a factor of $\sqrt{2}$. We, however, use the projector operator in Eq.\ \eqref{eq:EPRprojector} so that we do not have the factor $\sqrt{2}$ in all the equations below.

Then, if we denote $\psi_{\textrm{in}}(x)$ as a wave function of the input mode, the output after the conditional teleportation before the displacement operations can be written down as
\begin{multline}
\ket{\psi^\prime}=\hat{D}^\dagger_{p}(m_p)\hat{D}^\dagger_{x}(m_{x})\int\int\textrm{d}x\textrm{d}x^\prime\,e^{im_p(x-x^\prime)}\\
\psi_{\textrm{in}}(x)\Psi_\textrm{TMSS}(x-m_x,x^\prime-m_x)\ket{x^\prime}.
\end{multline}
In the normal context, by displacing, we can recover the input state and remove the dependence on the measurement results when $\Psi_\textrm{TMSS}(x,x^\prime)\to\delta(x-x^\prime)$.

As we have previously stated, since the $\ket{\textrm{NGES}(k,l)}$ is equivalent to $\ket{\textrm{TMSS}}$ acted on with $\hat{f}_{k,l}$, we can write down the (unnormalized) state when $\ket{\textrm{NGES}(k,l)}$ is used as
\begin{widetext}
\begin{align}
\ket{\psi^\prime}=\hat{f}_{k,l}(\eta)\hat{D}^\dagger_{p}(m_p)\hat{D}^\dagger_{x}(m_{x})\int\int\textrm{d}x\textrm{d}x^\prime\,e^{im_p(x-x^\prime)}\psi_{\textrm{in}}(x)\Psi_\textrm{TMSS}(x-m_x,x^\prime-m_x)\ket{x^\prime}.
\end{align}
\end{widetext}
Therefore, the wave function $\psi_\textrm{out}(x^\prime,m_x,m_p)$ after displacement operations is
\begin{multline}
\psi_\textrm{out}(x^\prime,m_x,m_p)=\\
\hat{h}_{k,l}(\eta,m_x,m_p)\psi_\textrm{cond}(x^\prime,m_x,m_p)
\end{multline}
where each part is defined as follows.
\begin{multline}
\hat{h}_{k,l}(\eta,m_x,m_p)\equiv\\
\hat{D}_{p}(m_p)\hat{D}_{x}(m_x)\hat{f}_{k,l}(\eta)\hat{D}^\dagger_{p}(m_p)\hat{D}^\dagger_{x}(m_{x})
\end{multline}
\begin{multline}
\psi_\textrm{cond}(x^\prime,m_x,m_p)\equiv\\
\int\textrm{d}x\,e^{im_p(x-x^\prime)}\psi_{\textrm{in}}(x)\Psi_\textrm{TMSS}(x-m_x,x^\prime-m_x)
\end{multline}

Let us consider the physical intuition of each part of $\psi_\textrm{out}(x^\prime,m_x,m_p)$. The conditional wave function $\psi_\textrm{cond}(x^\prime,m_x,m_p)$ consists of three parts: modulation due to $m_p$, input wave function $\psi_\textrm{in}(x)$, and convolution due to the TMSS. We observe that even if we displace both $x$ and $x^\prime$ of the $\Psi_\textrm{TMSS}(x,x^\prime)$, the argument of the first exponential term in Eq.\ \eqref{eq:TMSSwave} remains the same. Thus, the integral is evaluated around $x\approx x^\prime$. This makes the demodulation term with $m_p$ negligible as long as $m_p$ is not too big and the squeezing level is sufficient. On the other hand, the argument of the second exponential term in Eq.\ \eqref{eq:TMSSwave} is now centered around $m_x$. As such the effect due to the finite squeezing of the TMSS is the Gaussian convolution and a Gaussian envelope.

Next, regarding the term $\hat{h}_{k,l}(\eta,m_x,m_p)$, let us look at a special case where $k=1$, $l=0$, and $\eta\to 1$. In such case, this term becomes $\hat{h}_{k,l}(1,m_x,m_p)=\hat{x}-m_{x}$. Therefore, in the infinite squeezing limit, the wave function of the output state is equal to the wave function of the input state multiplied by $(x-m_x)$. 

Using this setup, it is possible to tailor the wave function of the input states. For example, if we consider $\ket{p=0}$ as our input state and iteratively use this circuit for the case where $k=1$ and $l=0$, then after $n$ iterations, the wave function becomes $\psi(x)=\prod_{i=1}^n(x-m_{i})$, where $m_i$ are the measurement results of the homodyne detectors. As we can see, $\psi(x)$ is a $n$-th order polynomial with real roots. We could also have imaginary roots by using, for example, $\hat{f}_{2,0}(\eta)$. In general, we can use this setup and realize arbitrary wave function with real roots. Although we believe that there should exist a modification of our setup to include arbitrary complex roots and complex coefficients, we leave the explorations of such possibility to the future work. Note that in a realistic setup, we would have to approximate $\ket{p=0}$ with $p$-squeezed states, which means that the wave function will be attenuated at large $x$. Moreover, the additional Gaussian envelope due to the finite squeezing and finite-width of the conditioning window $m_x$ and $m_p$ must be taken into an account.

In the next section, we will show the simulation results of various quantum states that can be generated with our system. We will first assume that the conditioning is applied with zero width. Afterward, the discussions on the actual conditioning window and success probability will be given.

\section{Simulations of the generated states\label{sec:simulation}}
In most of the simulations, we will be modest and assume the squeezing parameter to be $\vert r\vert\leq1.0$, which corresponds to about $-8.7$ dB of squeezing, an achievable value in the optical system \cite{PhysRevLett.117.110801}. Moreover, all calculations and simulations use quadrature basis and not Fock basis and the Wigner functions are calculated from the wave functions. For the cases where multiple iterations are used, we put $\mathbf{m}_x=(m_{x,1},m_{x,2},\dots,m_{x,n})$ and $\mathbf{m}_{p}=(m_{p,1},m_{p,2},\dots,m_{p,n})$ as vectors showing the values we condition the homodyne measurements in each iteration with.
\subsection{Schr{\"o}dinger cat states}
\begin{figure*}
\includegraphics[width=\textwidth]{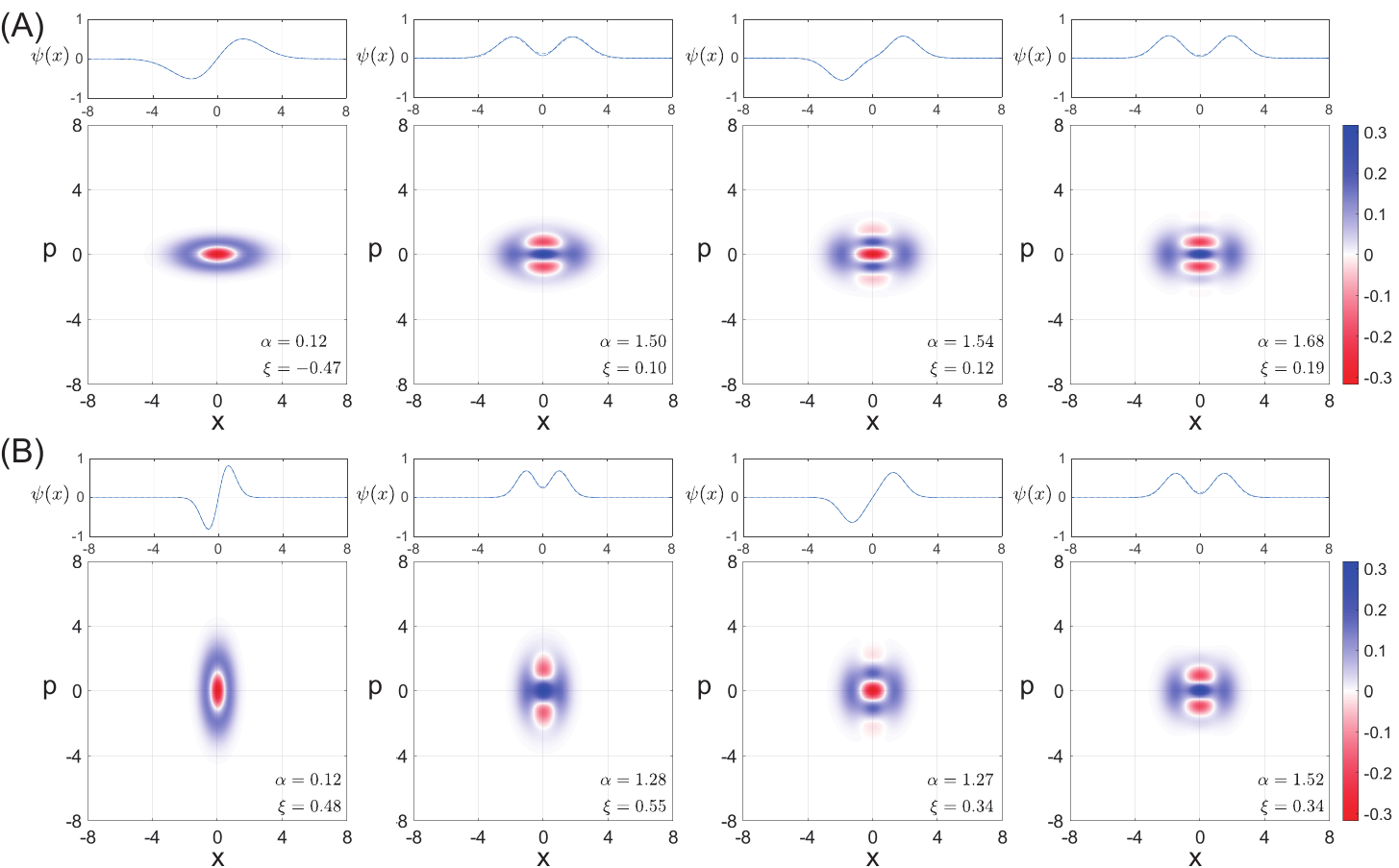}%
\caption{Simulation results of the wave functions and the Wigner functions of the generation of the Schr{\"o}dinger cat states. We use $k=1$ and $l=0$ in this simulation. The squeezing parameters for the NGES are $r_\textrm{tele}=1.0$ and the squeezing parameters $r$ of the initial squeezed states for (A) and (B) are $-1.0$ and $1.0$, respectively. Wave functions of the generated states (solid lines) and of the closest squeezed cat states $\hat{S}(\xi)\ket{\textrm{CAT},\alpha,\pm}$ (dashed lines) are shown. The parameters $\alpha$ and $\xi$ are written in the lower-right corner of each subfigure. The numbers of the iterations are $n=1,2,3,4$ (from left to right).\label{fig:cat_state}}
\end{figure*}

Here, we will show that our system can be used to generate cat states. A Schr{\"o}dinger cat state $\ket{\textrm{CAT},\alpha,\pm}$ is given by 
\begin{align}
\ket{\textrm{CAT},\alpha,\pm}=N_{\alpha,\pm}\left(\ket{\alpha}\pm\ket{-\alpha}\right),
\end{align}
where $N_{\alpha,\pm}$ is a normalization factor and we will assume $\alpha\in\mathbb{R}$ for the simplicity. We call $\ket{\textrm{CAT},\alpha,+}$ a plus cat state and $\ket{\textrm{CAT},\alpha,-}$ a minus cat state. In general, the generated states in this section will be close to the squeezed cat state $\hat{S}(\xi)\ket{\textrm{CAT},\alpha,\pm}$ and we will consider the fidelity of the generated states to the squeezed cat state with parameter $(\xi,\alpha)$.

Let us first consider the infinite squeezing limit. If we start from a $p$-squeezed as our input and implement $\hat{x}^{n}$ on it by repeating the circuit in Fig.\ \ref{fig:setup} for $n$ times, the (unnormalized) wave function in both $x$ and $p$ becomes
\begin{align}
\psi_{n}(x)&=x^{n}\exp\left(-\frac{x^2}{2e^{2r}}\right),\label{eq:cat_approx_x}\\
\tilde{\psi}_{n}(p)&=H_{n}\left(\frac{p}{\sqrt{2}e^{-r}}\right)\exp\left(-\frac{p^2}{2e^{-2r}}\right).\label{eq:cat_approx_p}
\end{align}
This type of wave function is similar to those in Ref. \cite{Ourjoumtsev2007,PhysRevA.103.013710} and is known to be a good approximation to the cat state. When $n$ is odd (even), the generated state approximates minus (plus) cat state. The wave function $\psi_{n}(x)$ has two extrema located at
\begin{align}
x_{\textrm{ext}}=\pm\sqrt{n}e^{r}
\end{align}
This means that the amplitude of the cat state will roughly scale with square root of the number of the iterations and the initial scale of the cat state will be determined by the squeezing of the input. In the $p$ quadrature, there is an oscillatory structure due to the Hermite polynomial. Note that in addition to $\hat{g}_{1,0}$, we can also consider using the NGES with more number of photon subtracted. In such a case, since $\hat{g}_{k,0}\propto H_{k}(i\hat{x})$, we expect that the increase in the amplitude will be amplified by roughly $\sqrt{k}$, since the leading term when $x$ becomes large is $\hat{x}^{k}$. Moreover, if we plot $H_{k,0}(ix)$, we can see that its values around $x=0$ are very small and the function rapidly increases at the large $x$. Such behavior is advantageous for cat state generations as we want the $\psi(x)$ to have two peaks that are far apart from each other.
 
In some protocols such as generation of GKP qubits using cat states, it is more advantageous to use squeezed cat states, rather than normal cat states with large amplitude \cite{PhysRevA.97.022341}. In the usual photon subtraction, a conventional method to approximate cat state, since the annihilation operators $\hat{a}$ are applied to the state, there can be no phase information and the cat states are always generated with the amplitude pointing out in the antisqueezing direction. In our method, since we are effectively applying quadrature operator $\hat{x}$, it is possible to apply $\hat{x}$ to a $x$-squeezed states so that the amplitudes is in the squeezing direction. The resulting wave functions after $n$ iterations in the infinite squeezing limit are
\begin{align}
\psi_{n}(x)&=x^{n}\exp\left(-\frac{x^2}{2e^{-2r}}\right),\\
\tilde{\psi}_{n}(p)&=H_{n}\left(\frac{p}{\sqrt{2}e^{r}}\right)\exp\left(-\frac{p^2}{2e^{2r}}\right),
\end{align}
which are simply the squeezed version of Eqs.\ \eqref{eq:cat_approx_x} and  \eqref{eq:cat_approx_p}. 

\begin{figure}
\includegraphics[width=\columnwidth]{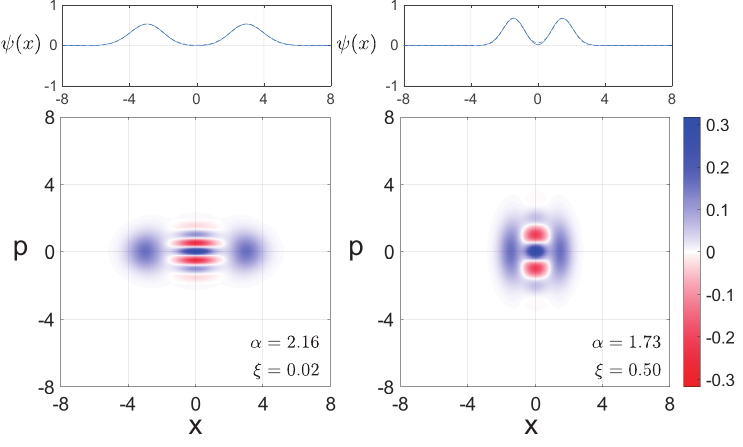}%
\caption{Simulation results of the generation of the Schr{\"o}dinger cat states for the case where $k=3$ and $l=0$. The squeezing parameters for the NGES are $r_\textrm{tele}=1.0$ and the squeezing parameters $r$ of the initial squeezed states are $-1.0$ and $1.0$ for the left and the right subfigure, respectively. The number of the iterations is 2.\label{fig:cat_state_big}}
\end{figure}

Figure \ref{fig:cat_state} shows the simulation results. We observe that the number of the interference fringes, which are characteristics of cat states, increases with each iteration. Contrary to our expectation, however, even for the case where we used $p$-squeezed states as our inputs, the resulting states are weakly squeezed in the $x$-direction. This is the effects from both the finite squeezing of the initial squeezed states and the finite squeezing in NGES. Even then, we observe that for the case where the initial state is squeezed in $\hat{x}$, the generated states is also squeezed in $\hat{x}$ without having to additionally squeeze the state. However, the amplitude $\alpha$ tends to be larger for the cases where $p$-squeezed states are used. Even so, the fidelity to the squeezed cat states are over $0.995$ for every subfigure. The procedure here can be repeated to achieve cat states with large amplitudes. Although we restrict ourselves to the case where $k=1$ and $l=0$ in Fig.\ \ref{fig:cat_state}, by increasing $k$, we could reach the large-amplitude cat states with much fewer iterations. Figure \ref{fig:cat_state_big} shows an example of this. Thus, depending on the number of the photon that we can resolve, the number of the iterations can be adjusted properly.

\subsection{Four-component cat states}
\begin{figure}
\includegraphics[width=\columnwidth]{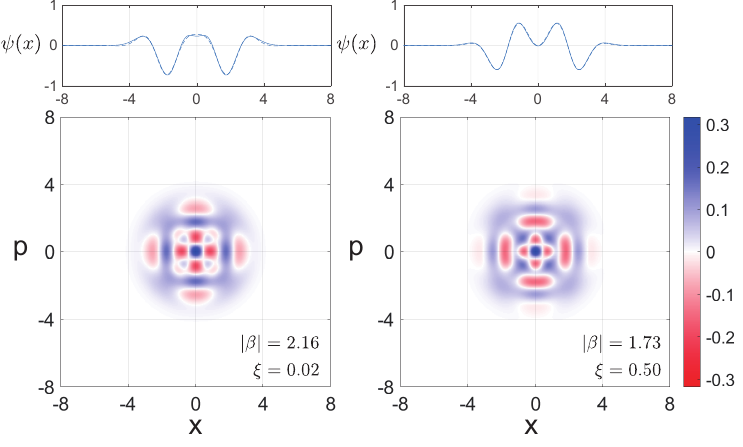}%
\caption{Four-component cat states generated with $\hat{f}_{1,1}$. The squeezing parameters for the NGES are $r_\textrm{tele}=1.0$ and the initial states are vacuum states. The amplitude of the closest four-component cat states are shown in the subfigure and the fidelity is above 0.99 for both subfigures. Note that these two are of different types of the four-component cat states. The numbers of the iterations are 3 and 4 for the left and the right subfigure, respectively.\label{fig:four_cat}}
\end{figure}
In addition to the usual cat state, there is also an important class of state called four-component cat state \cite{Hastrup:20,PhysRevLett.111.120501,PhysRevLett.119.030502}. This type of state is a superposition of four coherent states, i.e.\ $\ket{\psi_m}\propto\ket{\beta}+(-1)^m\ket{-\beta}+(i)^m\ket{i\beta}+(-i)^m\ket{-i\beta}$ with $\beta=\vert\beta\vert\exp(i\pi/4)$. Four-component cat states are known to be useful for applications such as cat codes \cite{PhysRevLett.111.120501,PhysRevLett.119.030502}. In this section, we will show that our system can also be used to generate such states.

Since four-component cat states are symmetric between $x$ and $p$, we should start with an input state with such symmetry. After that there are two ways where we can evolve the state into a four-component cat states. First, we can implement $\hat{f}_{1,0}$ and $\hat{f}_{0,1}$ alternatively with the conditioning at quadrature values 0 at the homodyne detectors. Note that this is also equivalent to keep implementing $\hat{f}_{1,0}$ but changing the measurement basis of the conditioning teleportation so that the teleported states are rotated by 90 degrees \cite{PhysRevA.90.062324}. Another method is to use $\hat{f}_{1,1}$ which is symmetric in $\hat{x}$ and $\hat{p}$ in the infinite squeezing limit. Note that we expect the operations to be symmetric in both quadrature for all $k=l$.

Figure \ref{fig:four_cat} shows the simulation results. We start with vacuum states and evolve the state in each iteration using $\hat{f}_{1,1}$. As $\hat{f}_{1,1}$ does not change the parity of the states, the generated states are four-component cat states whose wave functions are even functions. Although we only show the simulation results for the such cases, we can make the four-component cat states with odd wave functions by adding, for example $\hat{f}_{1,0}$.

\subsection{Fock state superpositions}
\begin{figure}
\includegraphics[width=\columnwidth]{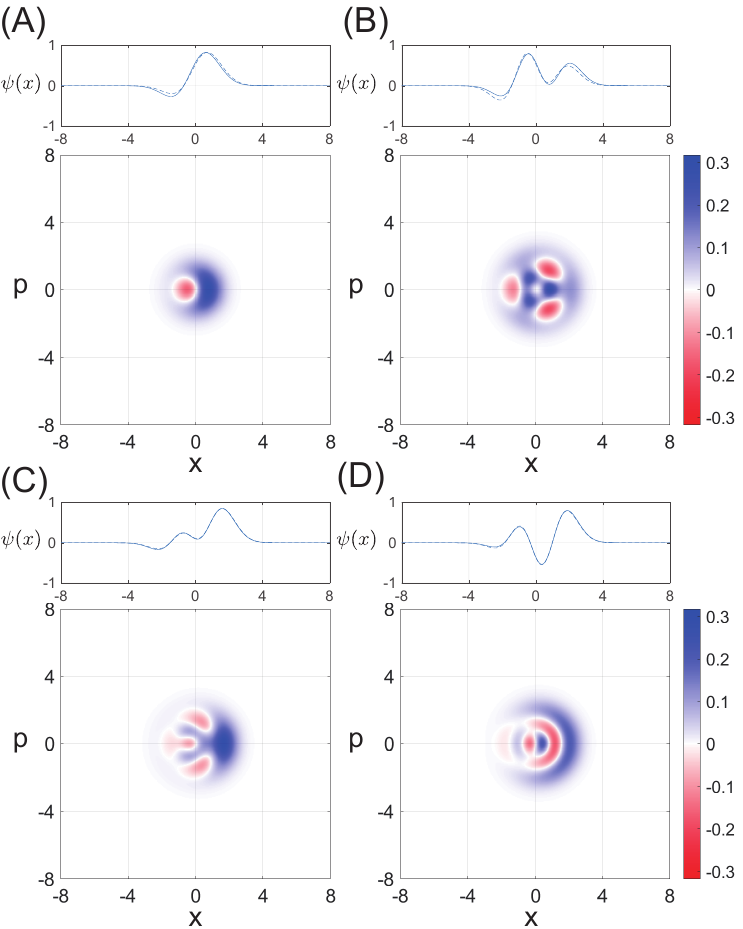}%
\caption{Simulation of  various superpositions of Fock states up to three photons. The squeezing parameters for the NGES are $r_\textrm{tele}=1.0$ and the initial states are vacuum states. (A) $\ket{0}+\ket{1}$. (B) $\ket{0}+\ket{3}$. (C) $\ket{0}+\ket{1}+\ket{2}+\ket{3}$. (D) $\ket{2}+\ket{3}$. For all cases, we assume $\mathbf{m}_{p}=\mathbf{0}$ and the $\mathbf{m}_{x}$ for each states are (-0.63), (-0.91, 0.93, 0.46), (-1,06, 0.13, 0.36), and (-1.27, 0.13, 0.99), respectively. The fidelity to the target states are 0.99, 0.97, $\sim$1.00, $\sim$1.00, respectively.\label{fig:Fock}}
\end{figure}

In the previous two examples regarding cat states, we assume that the results of the homodyne measurements are 0. In this section, to illustrate the possible applications of utilizing the other measurement results, we show how our system can be used to generate qubits. In the same way as a genuinely CV states can be approximated in Fock basis by truncating the infinitely large Hilbert spaces to subspaces below certain photon number, our method of tailoring wave function can also be used to approximate superposition of Fock states. This is done by attempting to shape the wave function to be close to the wave function of the Fock state superpositions.

For a Fock state superposition up to $n_\textrm{max}$ photons, the ket vector becomes
\begin{align}
\ket{\psi}=\sum_{n=0}^{n_\textrm{max}}c_{n}\ket{n}.
\end{align}
Then, the corresponding wave function is
\begin{align}
\ket{\psi}=\frac{1}{\pi^{1/4}}\sum_{n=0}^{n_\textrm{max}}c_{n}\frac{1}{\sqrt{2^n n!}}H_{n}(x)\exp\left(-\frac{x^2}{2}\right).
\end{align}
Therefore, for $c_{n}\in\mathbb{R}$, the wave function is a polynomial of at most $n_\textrm{max}$ order, multiplied with the wave function of the vacuum state. Therefore, we expect that atmost $n_\textrm{max}$ iterations of our circuit on vacuum states is required to generate such states. Note that we could also easily make the squeezed version of the Fock superposition state, $\hat{S}(r)\ket{\psi}$, by injecting squeezed states instead of the vacuum states and scaling the conditioning of the homodyne measurement results.

As an example, we consider Fock superposition states of the form $\ket{\psi_{F}}=c_{0}\ket{0}+c_{1}\ket{1}+c_{2}\ket{2}+c_{3}\ket{3}$. This type of arbitrary superposition of Fock states have been experimentally realized using TMSS, three avalanche photodiodes, and three coherent beams for displacements \cite{Yukawa:13}. When $c_{2}=c_{3}=0$, $\ket{\psi_{F}}$ reduces to the usual qubit. Figure \ref{fig:Fock} shows the simulation results. We observe that they have high fidelity to the target state and show how conditioning at other measurement results than 0 can be used to tailor the wave function. The general strategy here is that we select initial $\mathbf{m}_x$ so that they are close to the 0 points or the local minima of the wave functions and then optimize the wave function to match the target states. Another remark is that, unlike most experiment with Fock basis that usually implicitly assume low pump limit and truncate multiphoton components, no truncations in Fock basis are used and we work on the Hilbert space using quadrature basis.

\subsection{Approximated cubic phase state}
One of important classes of the nonclassical state is a cubic phase state $\ket{\textrm{CPS}}=\exp(i\gamma\hat{x}^3)\ket{p=0}$. This state is an ancillary state for cubic phase gate which is a candidate for non-Gaussian gates in CV quantum computation \cite{PhysRevA.64.012310,PhysRevA.93.022301}. In the previous attempts to generate this state, truncation based on Fock state was used and this state was approximated to up to three photons \cite{Yukawa:13}. In this section, we will show that it is possible to use our method to realize cubic phase state.

If we expand the ket vector of the cubic phase state, we get
\begin{align}
\ket{\textrm{CPS}}=\left[1+i\gamma\hat{x}^3+\frac{1}{2!}(i\gamma\hat{x}^3)^2+\dots\right]\ket{p=0}.
\end{align}
As it was previously mentioned in Sec.\ \ref{subsec:output}, our method can realize wave function in $x$ that has real roots. Therefore, we need to modified the above equation. By looking at the wave function in $p$ instead of $x$ and approximating $\ket{p=0}$ with $p$-squeezed states, we get
\begin{widetext}
\begin{align}
\tilde{\psi}_{\ket{\textrm{CPS}}}(p)\approx\left[1+\gamma^\prime H_{3}\left(\frac{p^\prime}{\sqrt{2}}\right)+\frac{\gamma^{\prime2}}{2!}H_{6}\left(\frac{p^\prime}{\sqrt{2}}\right)+\dots\right]\exp\left(-\frac{p^{\prime 2}}{2}\right),\label{eq:cps_photon}
\end{align}
\end{widetext}
with $\gamma^{\prime}=\gamma/(\sqrt{2}e^{-2\xi})^3$ and $p^\prime=p/e^{-\xi}$. In that sense, the parameter $\gamma$ and the squeezing of the squeezed states are in the scaling relationship. In the approximation, we will assume that $\xi=0$ for simplicity, as the squeezing can be added after the state is generated when necessary.

There is also another possible approximation. If we recall that the unnormalized wave function of the ideal $\ket{\textrm{CPS}}$ is
\begin{align}
\psi_{\ket{\textrm{CPS}}}(x)\propto\exp(i\gamma\hat{x}^3),
\end{align}
the Fourier transform of the above function is \cite{doi:10.1080/09500340601101575}
\begin{align}
\tilde{\psi}_{\ket{\textrm{CPS}}}(p)\propto\Ai\left(-\frac{p}{\sqrt[3]{3\gamma}}\right),\label{eq:airy}
\end{align}
where $\Ai(\cdot)$ is the Airy function. As the Airy function is a real function, it can be approximated using our methodology. However, there are two points that need considerations. First, as $\tilde{\psi}_{\ket{\textrm{CPS}}}(p)$ is mainly contained in the upper-region of the phase space (i.e.\ region with positive $p$), it will be more advantageous to generate a displaced version of this. Second, ideal Airy functions extend to the infinity, meaning that we have to add another approximation. One of possible ways to do so is considering a Gaussian envelope over a displaced Airy function as our approximation, i.e.
\begin{align}
\tilde{\psi}_{\ket{\textrm{CPS}}}(p)\approx\exp\left(-\frac{p^2}{2e^{2\xi}}\right)\Ai\left(-\frac{(p+p_0)}{\sqrt[3]{3\gamma}}\right)\label{eq:airy_2}
\end{align}
as our target state which approaches ideal $\ket{\textrm{CPS}}$ in the limit of $\xi\to\infty$.

\begin{figure}
\includegraphics[width=\columnwidth]{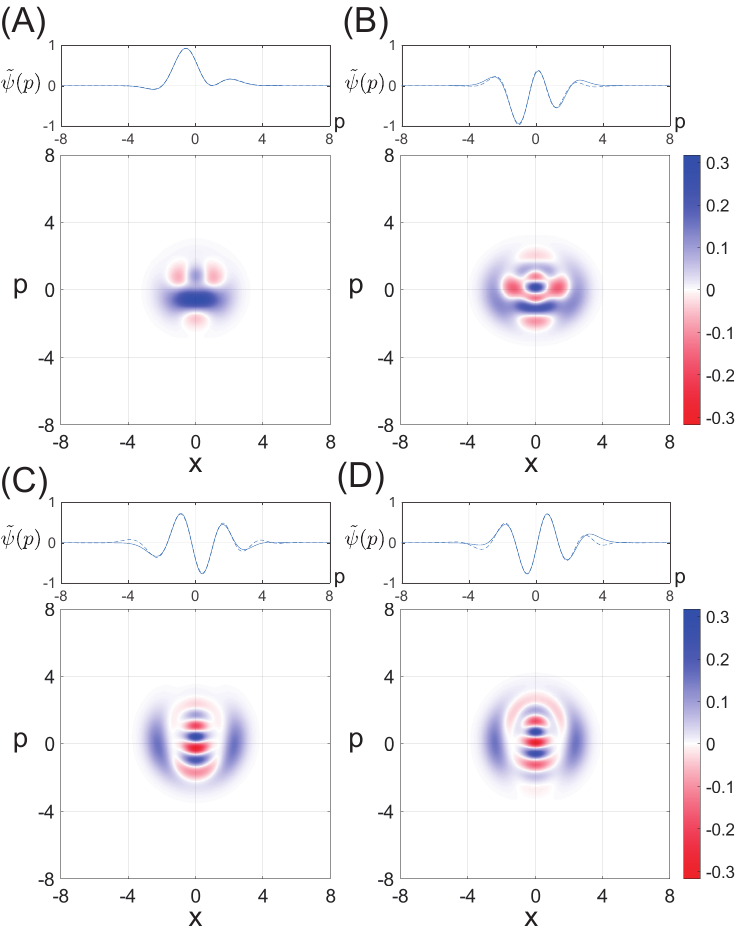}%
\caption{Simulations of cubic phase state generations. The target states for all subfigures assume $\gamma=0.5$ and $\mathbf{m}_p=\mathbf{0}$ and generation using $\hat{f}_{1,0}$. The squeezing parameters of the NGES for all subfigures are $r_\textrm{tele}=1.0$. (A,B) Approximation using Eq.\ \eqref{eq:cps_photon} with $\xi=0$ up to the first-order (A) and the second-order (B) in $\gamma$. The initial squeezing level of the inputs is $r=0$ and $-0.7$, and the $\mathbf{m}_{x}$ are (0.78, -1.51, 0.58) and (0.61, -1.15, -0.23, 0.60), respectively. (C,D) Approximation using Eq.\ \eqref{eq:airy_2}. The squeezing of the envelope is $\xi=0.6$ and $p_0=8$ and $9$, respectively. The $\mathbf{m}_x$ are (2.80, 1.39, -1.18, -0.08, 1.02) and (-1.73, 1.72, -0.68, 1.02, 0.08). The fidelities of each state to its targeted approximated $\ket{\textrm{CPS}}$ are $\sim 1.00$, 0.985, 0.978, and 0.962, respectively. Note that the state is rotated by 90 degrees after the generation. \label{fig:CPS}  }
\end{figure}
Figure \ref{fig:CPS} shows the generated state using both approximation that is targeted to this state. We observe that although both approximations yield different Wigner functions, they share similar traits: parabolic structure and oscillatory structure in $p$-direction. The first type of approximation is actually equivalent to Fock basis truncations when $\xi=0$ is equivalent to truncation up to six photons. On the other hand, the second type of approximation would roughly be equivalent to reducing the weight of the multiphoton components, but not completely truncating them in a sense that the Wigner functions of the Fock states with lower photon numbers tend to be localized near the origin of the phase space. Note that a more rigorous approximation of the cubic phase state is to look at its performance when it is used to realize the cubic phase gate \cite{PhysRevA.93.022301,2020arXiv201114576K}.

\section{Discussions\label{sec:discussion}}

\subsection{Success rate and fidelity}

In this section, we discuss success rate and fidelity. Up until this point, we have assumed for the simplicity that the window of the conditioning can be infinitesimal. In reality, we have to have a finite window to have a finite success rate. Since we are effectively tailoring by approximating them as polynomial and the measurement results of the homodyne detector determine the positions of the roots of the polynomials, the conditioning window size is determined by how the polynomials change when the roots are changed. In general, we would expect that for a wave function that is broadly distributed, the positions of the roots do not greatly affect the overall polynomials, thus we can have a large conditioning window. This means that, it tends to be more advantageous to tailor the wave function of the antisqueezed version of the target state.

\begin{figure}
\includegraphics[width=\columnwidth]{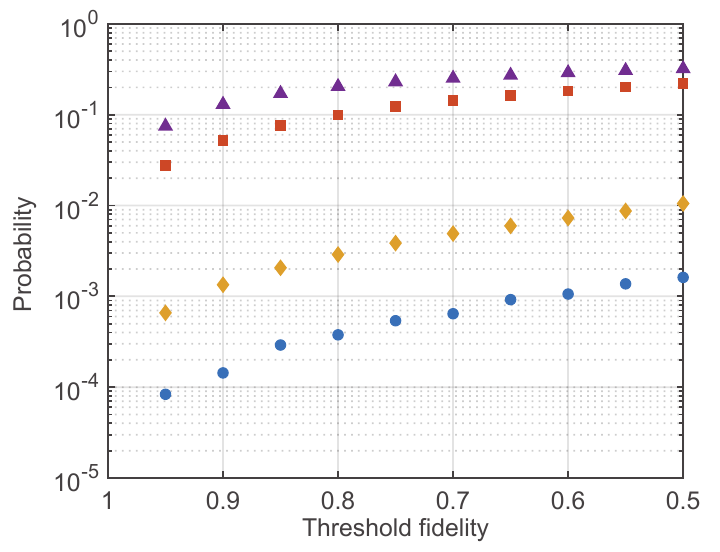}%
\caption{Success probabilities of the homodyne conditioning for states generated with a single step of $\hat{f}_{1,0}$ to have fidelities to the target states above the threshold fidelity. The initial states are $\hat{S}(r)\ket{0}$ (circle), $\hat{S}(-r)\ket{0}$ (diamond), $\hat{S}(r)\ket{1}$ (square), and $\hat{S}(-r)\ket{1}$ (triangle), with $r=1.0$ and $r_\textrm{tele}=1.0$. The target states are the states generated when $m_x=m_p=0$ for all input states.\label{fig:probability}}
\end{figure}

Figure \ref{fig:probability} shows the probability of successfully generating quantum states whose fidelity to the target state is higher than a certain threshold. Indeed, the success probability is higher as we lower the threshold fidelity. We also observe that, as expected, the success probability is indeed higher for a state that is broadly distributed in the quadrature $x$ for the current case where the operator $\hat{f}_{1,0}$ is applied to the initial state in the case where the conditioning window is infinitesimal. Interestingly, although we are conditioning near $m_x=m_p=0$, rather than squeezed states whose wave function is a Gaussian function centered at the origin, the squeezed single photon states, which have zero probability of finding the quadrature value 0, have much higher success probability. This is due to the fact that when the squeezed single photon is interfered with one of the modes of the $\ket{\textrm{NGES}(1,0)}$, the probability of Bell measurement giving $m_x=m_p=0$ increases due to the interference of the non-Gaussian features of both states.

For the case where the conditioning teleportation with NGES is implemented multiple times, there are additional aspects that must be considered. First, the overall success probability. To illustrate this aspect quantitatively, let us consider the ideal case with $k=1$ and $l=0$ and infinite squeezing. Then, if the initial wave function is $\psi(x)$, the unnormalized wave function after one and two step of our circuit is $(x-m)\psi(x)$ and $(x-m^{\prime})(x-m^{\prime\prime})\psi(x)$ where $m$, $m^\prime$, and $m^{\prime\prime}$ are the results of the Bell measurement in $x$ quadrature. If we want to do conditioning near the place where $m=m^\prime=m^{\prime\prime}=0$, then for the one-step case, the allowable range of $m$ should be well below $\langle\hat{x}^2\rangle^{1/2}$, where the mean $\langle\cdot\rangle$ here is taken with respect to the initial state $\ket{\psi}$. On the other hand, for the two-step case, if we restrict to the case where $m^\prime=-m^{\prime\prime}$, it is easy to show that the allowable range should be well below $\langle\hat{x}^4\rangle^{1/4}$. As $\langle\hat{x}^4\rangle^{1/4}\geq\langle\hat{x}^2\rangle^{1/2}$, the allowable range of each measurement in the two-step case should be broader given that they have appropriate relation. As such, this qualitative example suggests that for the multi-step case, rather than consider each step individually, we should perform conditioning and heralding on a set or a range of the measurement results.

\begin{figure}
\includegraphics[width=\columnwidth]{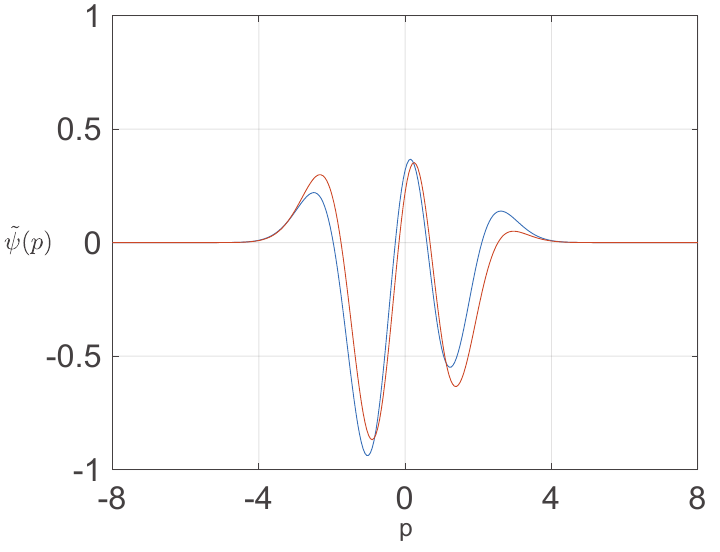}%
\caption{Effects due to the shifts of the measurement results. Blue line: The wave function of the approximated cubic phase state in Fig.\ \ref{fig:CPS}(B). Orange line: the wave function of the state generated when the measurement results $\mathbf{m}_x$ is shifted by (0.1, 0.1, 0.1, 0.1).\label{fig:displaced}}
\end{figure}
Another aspect we have to consider is the possibility of increasing the fidelity to the target state using Gaussian operations. Figure \ref{fig:displaced} shows an example of such cases. Although the fidelity between the two wave functions is 0.87, it is obvious that the two are related via displacement operation. If the wave function is displaced by the amount of the shifts, the fidelity becomes 0.95 which is much higher. The effects from the displacements due to the shifts in the measurement results from the target state are more obvious when the wave function is oscillatory as shown in this example. In addition, we could also consider a case where all the measurement results are just the scaling of the target case. In such case, the generated state would be roughly the squeezed or antisqueezed version of the target state. Therefore, when considering the success probability and the fidelity, it is more advantageous to optimize the fidelity to the target state using Gaussian operations. This optimization is expected to increase the success probability further. 

By combining these two aspects, it is expected that the success probability of our method can be further increased. As the calculations for the actual experimental setup would be highly dependent on the initial states and the target states, and also the experimental imperfections such as optical losses, we leave the detailed consideration as a future experimental work.

\subsection{Experimental feasibility}
In addition to the probabilistic nature of the homodyne conditioning, we also need to consider how to realize NGES. We will restrict our discussions to the realization in the optical systems. One of the simplest implementations to realize NGES is that via photon subtraction, which is a method widely used to approximate cat states. As photon subtraction is probabilistic, using the photon subtraction as it is will limit the generation rate even further. There are a few possibilities to overcome this. First, we could simply employ quantum memory as our protocol used NGES as resource states that can be generated offline. Second, for $k=1$ and $l=0$ (or $k=0$ and $l=1$), the squeezed states become squeezed single photon states. As there exists on-demand single photon source based on architecture such as quantum dots \cite{doi:10.1063/5.0010193} and deterministic squeezing of single photons have been realized \cite{PhysRevLett.113.013601}, we could use such architecture for deterministic generation. Third, as photon-subtracted squeezed states are usually approximations of cat states \cite{PhysRevA.55.3184}, we could also consider a possibility of replacing them with cat state sources. Recently, there is a proposal for a system to generate  Schr{\"o}dinger cat state with high generation rate \cite{PhysRevA.103.013710} which might enable realistic realization of the method in this paper.

\subsection{Time-domain-multiplexing and Non-Gaussian cluster states}
\begin{figure}
\includegraphics[width=\columnwidth]{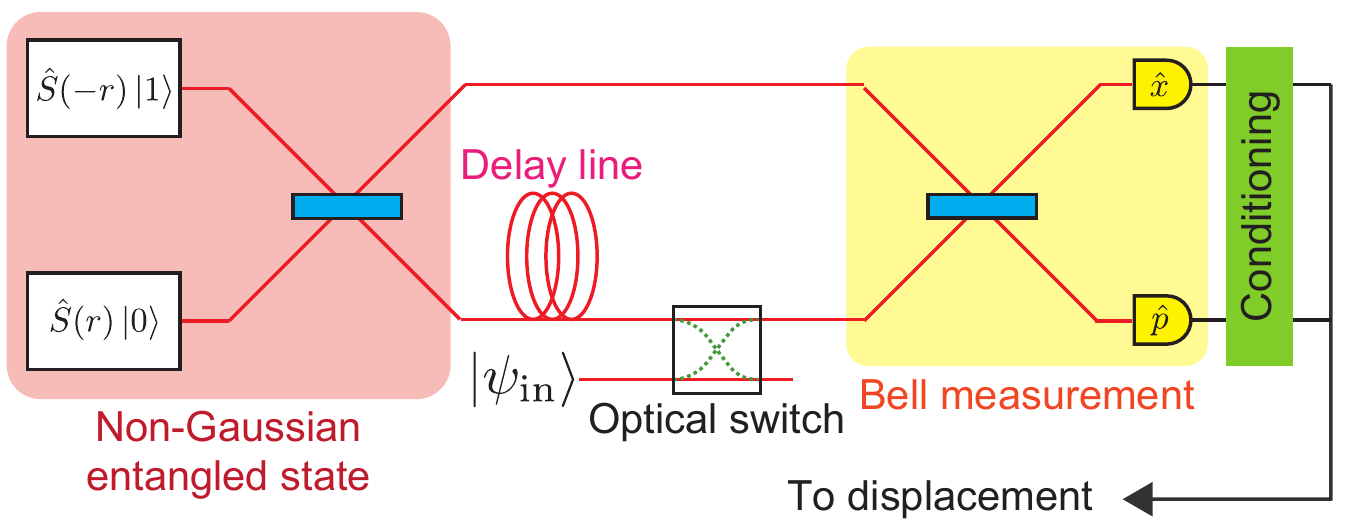}%
\caption{A setup for iterative implementation of our protocol using the time-domain multiplexing method. The NGES used is for the implementation of $\hat{f}_{1,0}$ and is realized by replacing one of the squeezed light source in the one-dimensional cluster state setup with a squeezed single photon source.\label{fig:TDM_setup}}
\end{figure}
Figure \ref{fig:TDM_setup} shows a possible experimental realization of our setup using the time-domain-multiplexing method. As our setup is based on sequential quantum teleportation circuits, by using the time-domain multiplexing and replacing the source of the TMSS with the NGES, it is possible to realize a versatile and compact experimental setup for generation of various non-Gaussian states, where the generated state can be programmed via the choice of the homodyne conditioning. The optical switch is used to inject the initial quantum state or retrieve the quantum state after the implementation of our protocol. The usage of such optical switch on the quantum states has recently been demonstrated \cite{Takedaeaaw4530,Larsen2019}.

Also, as the setup in Fig.\ \ref{fig:TDM_setup} resembles that of the one-dimensional cluster state generation \cite{PhysRevA.83.062314,doi:10.1063/1.4962732,Yokoyama2013}, we could also interpret our protocol as a cluster state computation where the NGES is used instead of the usual Gaussian CV cluster states. Degaussification of Gaussian cluster states and their properties have been studied recently \cite{PhysRevLett.121.220501,PhysRevLett.124.150501,PRXQuantum.1.020305} and a basic demonstration of the generation of such non-Gaussian cluster state has also been realized \cite{Ra2020}. In this context, our protocol here demonstrates a possible application of non-Gaussian entanglement resource and could serve as a setup to study the combination of the non-Gaussian element to the time-domain-multiplexed cluster state.

\section{Conclusion\label{sec:conclusion}}
In this paper, we present a methodology to generate non-Gaussian states based on tailoring of the wave function using the non-Gaussian entanglements and the conditional quantum teleportation. Our approach is a complementary approach to the Fock basis approach. We demonstrate the versatility of our method by showing that, without modifying our system at all, we can generate various quantum states with iterative conditional quantum teleportation using the non-Gaussian entanglement resources. In addition to a system that can generate any states, we could also consider a system specialized for a certain state such as GKP states. Such a study has been done in Ref.\ \cite{PhysRevA.101.032315} where parameters of the optical systems and photon number resolving detector are optimized in Fock basis. As we have demonstrated that homodyne conditioning is also useful for state generation, it will be interesting to see if we can also incorporate homodyne conditioning and realize a kind of hybridized state generator that utilizes both Fock basis and wave function picture. 

\begin{acknowledgments}
This work was partly supported by JST [Moonshot R\&D][Grant No.\ JPMJMS2064], JSPS KAKENHI (Grant No.\ 18H05207, No.\ 18H01149, and No.\ 20K15187), UTokyo Foundation, and donations from Nichia Corporation. K.T. acknowledges financial supports from the Japan Society for the Promotion of Science (JSPS). The authors would like to thank Takahiro Mitani for proofreading of the manuscript.
\end{acknowledgments}

\bibliography{wavefunction_engineering_ref.bib}

\begin{thebibliography}{62}%
\makeatletter
\providecommand \@ifxundefined [1]{%
 \@ifx{#1\undefined}
}%
\providecommand \@ifnum [1]{%
 \ifnum #1\expandafter \@firstoftwo
 \else \expandafter \@secondoftwo
 \fi
}%
\providecommand \@ifx [1]{%
 \ifx #1\expandafter \@firstoftwo
 \else \expandafter \@secondoftwo
 \fi
}%
\providecommand \natexlab [1]{#1}%
\providecommand \enquote  [1]{``#1''}%
\providecommand \bibnamefont  [1]{#1}%
\providecommand \bibfnamefont [1]{#1}%
\providecommand \citenamefont [1]{#1}%
\providecommand \href@noop [0]{\@secondoftwo}%
\providecommand \href [0]{\begingroup \@sanitize@url \@href}%
\providecommand \@href[1]{\@@startlink{#1}\@@href}%
\providecommand \@@href[1]{\endgroup#1\@@endlink}%
\providecommand \@sanitize@url [0]{\catcode `\\12\catcode `\$12\catcode
  `\&12\catcode `\#12\catcode `\^12\catcode `\_12\catcode `\%12\relax}%
\providecommand \@@startlink[1]{}%
\providecommand \@@endlink[0]{}%
\providecommand \url  [0]{\begingroup\@sanitize@url \@url }%
\providecommand \@url [1]{\endgroup\@href {#1}{\urlprefix }}%
\providecommand \urlprefix  [0]{URL }%
\providecommand \Eprint [0]{\href }%
\providecommand \doibase [0]{https://doi.org/}%
\providecommand \selectlanguage [0]{\@gobble}%
\providecommand \bibinfo  [0]{\@secondoftwo}%
\providecommand \bibfield  [0]{\@secondoftwo}%
\providecommand \translation [1]{[#1]}%
\providecommand \BibitemOpen [0]{}%
\providecommand \bibitemStop [0]{}%
\providecommand \bibitemNoStop [0]{.\EOS\space}%
\providecommand \EOS [0]{\spacefactor3000\relax}%
\providecommand \BibitemShut  [1]{\csname bibitem#1\endcsname}%
\let\auto@bib@innerbib\@empty
\bibitem [{\citenamefont {Nielsen}\ and\ \citenamefont
  {Chuang}(2000)}]{nielsen00}%
  \BibitemOpen
  \bibfield  {author} {\bibinfo {author} {\bibfnamefont {M.~A.}\ \bibnamefont
  {Nielsen}}\ and\ \bibinfo {author} {\bibfnamefont {I.~L.}\ \bibnamefont
  {Chuang}},\ }\href@noop {} {\emph {\bibinfo {title} {Quantum Computation and
  Quantum Information}}}\ (\bibinfo  {publisher} {Cambridge University Press},\
  \bibinfo {address} {Cambridge},\ \bibinfo {year} {2000})\BibitemShut
  {NoStop}%
\bibitem [{\citenamefont {Raussendorf}\ and\ \citenamefont
  {Briegel}(2001)}]{PhysRevLett.86.5188}%
  \BibitemOpen
  \bibfield  {author} {\bibinfo {author} {\bibfnamefont {R.}~\bibnamefont
  {Raussendorf}}\ and\ \bibinfo {author} {\bibfnamefont {H.~J.}\ \bibnamefont
  {Briegel}},\ }\bibfield  {title} {\bibinfo {title} {A one-way quantum
  computer},\ }\href {https://doi.org/10.1103/PhysRevLett.86.5188} {\bibfield
  {journal} {\bibinfo  {journal} {Phys. Rev. Lett.}\ }\textbf {\bibinfo
  {volume} {86}},\ \bibinfo {pages} {5188} (\bibinfo {year}
  {2001})}\BibitemShut {NoStop}%
\bibitem [{\citenamefont {Menicucci}\ \emph {et~al.}(2006)\citenamefont
  {Menicucci}, \citenamefont {van Loock}, \citenamefont {Gu}, \citenamefont
  {Weedbrook}, \citenamefont {Ralph},\ and\ \citenamefont
  {Nielsen}}]{PhysRevLett.97.110501}%
  \BibitemOpen
  \bibfield  {author} {\bibinfo {author} {\bibfnamefont {N.~C.}\ \bibnamefont
  {Menicucci}}, \bibinfo {author} {\bibfnamefont {P.}~\bibnamefont {van
  Loock}}, \bibinfo {author} {\bibfnamefont {M.}~\bibnamefont {Gu}}, \bibinfo
  {author} {\bibfnamefont {C.}~\bibnamefont {Weedbrook}}, \bibinfo {author}
  {\bibfnamefont {T.~C.}\ \bibnamefont {Ralph}},\ and\ \bibinfo {author}
  {\bibfnamefont {M.~A.}\ \bibnamefont {Nielsen}},\ }\bibfield  {title}
  {\bibinfo {title} {Universal quantum computation with continuous-variable
  cluster states},\ }\href {https://doi.org/10.1103/PhysRevLett.97.110501}
  {\bibfield  {journal} {\bibinfo  {journal} {Phys. Rev. Lett.}\ }\textbf
  {\bibinfo {volume} {97}},\ \bibinfo {pages} {110501} (\bibinfo {year}
  {2006})}\BibitemShut {NoStop}%
\bibitem [{\citenamefont {Yokoyama}\ \emph {et~al.}(2013)\citenamefont
  {Yokoyama}, \citenamefont {Ukai}, \citenamefont {Armstrong}, \citenamefont
  {Sornphiphatphong}, \citenamefont {Kaji}, \citenamefont {Suzuki},
  \citenamefont {Yoshikawa}, \citenamefont {Yonezawa}, \citenamefont
  {Menicucci},\ and\ \citenamefont {Furusawa}}]{Yokoyama2013}%
  \BibitemOpen
  \bibfield  {author} {\bibinfo {author} {\bibfnamefont {S.}~\bibnamefont
  {Yokoyama}}, \bibinfo {author} {\bibfnamefont {R.}~\bibnamefont {Ukai}},
  \bibinfo {author} {\bibfnamefont {S.~C.}\ \bibnamefont {Armstrong}}, \bibinfo
  {author} {\bibfnamefont {C.}~\bibnamefont {Sornphiphatphong}}, \bibinfo
  {author} {\bibfnamefont {T.}~\bibnamefont {Kaji}}, \bibinfo {author}
  {\bibfnamefont {S.}~\bibnamefont {Suzuki}}, \bibinfo {author} {\bibfnamefont
  {J.}~\bibnamefont {Yoshikawa}}, \bibinfo {author} {\bibfnamefont
  {H.}~\bibnamefont {Yonezawa}}, \bibinfo {author} {\bibfnamefont {N.~C.}\
  \bibnamefont {Menicucci}},\ and\ \bibinfo {author} {\bibfnamefont
  {A.}~\bibnamefont {Furusawa}},\ }\bibfield  {title} {\bibinfo {title}
  {Ultra-large-scale continuous-variable cluster states multiplexed in the time
  domain},\ }\href {https://doi.org/10.1038/nphoton.2013.287} {\bibfield
  {journal} {\bibinfo  {journal} {Nature Photonics}\ }\textbf {\bibinfo
  {volume} {7}},\ \bibinfo {pages} {982} (\bibinfo {year} {2013})}\BibitemShut
  {NoStop}%
\bibitem [{\citenamefont {Yoshikawa}\ \emph {et~al.}(2016)\citenamefont
  {Yoshikawa}, \citenamefont {Yokoyama}, \citenamefont {Kaji}, \citenamefont
  {Sornphiphatphong}, \citenamefont {Shiozawa}, \citenamefont {Makino},\ and\
  \citenamefont {Furusawa}}]{doi:10.1063/1.4962732}%
  \BibitemOpen
  \bibfield  {author} {\bibinfo {author} {\bibfnamefont {J.}~\bibnamefont
  {Yoshikawa}}, \bibinfo {author} {\bibfnamefont {S.}~\bibnamefont {Yokoyama}},
  \bibinfo {author} {\bibfnamefont {T.}~\bibnamefont {Kaji}}, \bibinfo {author}
  {\bibfnamefont {C.}~\bibnamefont {Sornphiphatphong}}, \bibinfo {author}
  {\bibfnamefont {Y.}~\bibnamefont {Shiozawa}}, \bibinfo {author}
  {\bibfnamefont {K.}~\bibnamefont {Makino}},\ and\ \bibinfo {author}
  {\bibfnamefont {A.}~\bibnamefont {Furusawa}},\ }\bibfield  {title} {\bibinfo
  {title} {Invited article: Generation of one-million-mode continuous-variable
  cluster state by unlimited time-domain multiplexing},\ }\href
  {https://doi.org/10.1063/1.4962732} {\bibfield  {journal} {\bibinfo
  {journal} {APL Photonics}\ }\textbf {\bibinfo {volume} {1}},\ \bibinfo
  {pages} {060801} (\bibinfo {year} {2016})},\ \Eprint
  {https://arxiv.org/abs/https://doi.org/10.1063/1.4962732}
  {https://doi.org/10.1063/1.4962732} \BibitemShut {NoStop}%
\bibitem [{\citenamefont {Asavanant}\ \emph {et~al.}(2019)\citenamefont
  {Asavanant}, \citenamefont {Shiozawa}, \citenamefont {Yokoyama},
  \citenamefont {Charoensombutamon}, \citenamefont {Emura}, \citenamefont
  {Alexander}, \citenamefont {Takeda}, \citenamefont {Yoshikawa}, \citenamefont
  {Menicucci}, \citenamefont {Yonezawa},\ and\ \citenamefont
  {Furusawa}}]{Asavanant373}%
  \BibitemOpen
  \bibfield  {author} {\bibinfo {author} {\bibfnamefont {W.}~\bibnamefont
  {Asavanant}}, \bibinfo {author} {\bibfnamefont {Y.}~\bibnamefont {Shiozawa}},
  \bibinfo {author} {\bibfnamefont {S.}~\bibnamefont {Yokoyama}}, \bibinfo
  {author} {\bibfnamefont {B.}~\bibnamefont {Charoensombutamon}}, \bibinfo
  {author} {\bibfnamefont {H.}~\bibnamefont {Emura}}, \bibinfo {author}
  {\bibfnamefont {R.~N.}\ \bibnamefont {Alexander}}, \bibinfo {author}
  {\bibfnamefont {S.}~\bibnamefont {Takeda}}, \bibinfo {author} {\bibfnamefont
  {J.}~\bibnamefont {Yoshikawa}}, \bibinfo {author} {\bibfnamefont {N.~C.}\
  \bibnamefont {Menicucci}}, \bibinfo {author} {\bibfnamefont {H.}~\bibnamefont
  {Yonezawa}},\ and\ \bibinfo {author} {\bibfnamefont {A.}~\bibnamefont
  {Furusawa}},\ }\bibfield  {title} {\bibinfo {title} {Generation of
  time-domain-multiplexed two-dimensional cluster state},\ }\href
  {https://doi.org/10.1126/science.aay2645} {\bibfield  {journal} {\bibinfo
  {journal} {Science}\ }\textbf {\bibinfo {volume} {366}},\ \bibinfo {pages}
  {373} (\bibinfo {year} {2019})},\ \Eprint
  {https://arxiv.org/abs/https://science.sciencemag.org/content/366/6463/373.full.pdf}
  {https://science.sciencemag.org/content/366/6463/373.full.pdf} \BibitemShut
  {NoStop}%
\bibitem [{\citenamefont {Larsen}\ \emph
  {et~al.}(2019{\natexlab{a}})\citenamefont {Larsen}, \citenamefont {Guo},
  \citenamefont {Breum}, \citenamefont {Neergaard-Nielsen},\ and\ \citenamefont
  {Andersen}}]{Larsen369}%
  \BibitemOpen
  \bibfield  {author} {\bibinfo {author} {\bibfnamefont {M.~V.}\ \bibnamefont
  {Larsen}}, \bibinfo {author} {\bibfnamefont {X.}~\bibnamefont {Guo}},
  \bibinfo {author} {\bibfnamefont {C.~R.}\ \bibnamefont {Breum}}, \bibinfo
  {author} {\bibfnamefont {J.~S.}\ \bibnamefont {Neergaard-Nielsen}},\ and\
  \bibinfo {author} {\bibfnamefont {U.~L.}\ \bibnamefont {Andersen}},\
  }\bibfield  {title} {\bibinfo {title} {Deterministic generation of a
  two-dimensional cluster state},\ }\href
  {https://doi.org/10.1126/science.aay4354} {\bibfield  {journal} {\bibinfo
  {journal} {Science}\ }\textbf {\bibinfo {volume} {366}},\ \bibinfo {pages}
  {369} (\bibinfo {year} {2019}{\natexlab{a}})},\ \Eprint
  {https://arxiv.org/abs/https://science.sciencemag.org/content/366/6463/369.full.pdf}
  {https://science.sciencemag.org/content/366/6463/369.full.pdf} \BibitemShut
  {NoStop}%
\bibitem [{\citenamefont {Chen}\ \emph {et~al.}(2014)\citenamefont {Chen},
  \citenamefont {Menicucci},\ and\ \citenamefont
  {Pfister}}]{PhysRevLett.112.120505}%
  \BibitemOpen
  \bibfield  {author} {\bibinfo {author} {\bibfnamefont {M.}~\bibnamefont
  {Chen}}, \bibinfo {author} {\bibfnamefont {N.~C.}\ \bibnamefont
  {Menicucci}},\ and\ \bibinfo {author} {\bibfnamefont {O.}~\bibnamefont
  {Pfister}},\ }\bibfield  {title} {\bibinfo {title} {Experimental realization
  of multipartite entanglement of 60 modes of a quantum optical frequency
  comb},\ }\href {https://doi.org/10.1103/PhysRevLett.112.120505} {\bibfield
  {journal} {\bibinfo  {journal} {Phys. Rev. Lett.}\ }\textbf {\bibinfo
  {volume} {112}},\ \bibinfo {pages} {120505} (\bibinfo {year}
  {2014})}\BibitemShut {NoStop}%
\bibitem [{\citenamefont {{Asavanant}}\ \emph {et~al.}(2020)\citenamefont
  {{Asavanant}}, \citenamefont {{Charoensombutamon}}, \citenamefont
  {{Yokoyama}}, \citenamefont {{Ebihara}}, \citenamefont {{Nakamura}},
  \citenamefont {{Alexander}}, \citenamefont {{Endo}}, \citenamefont
  {{Yoshikawa}}, \citenamefont {{Menicucci}}, \citenamefont {{Yonezawa}},\ and\
  \citenamefont {{Furusawa}}}]{2020arXiv200611537A}%
  \BibitemOpen
  \bibfield  {author} {\bibinfo {author} {\bibfnamefont {W.}~\bibnamefont
  {{Asavanant}}}, \bibinfo {author} {\bibfnamefont {B.}~\bibnamefont
  {{Charoensombutamon}}}, \bibinfo {author} {\bibfnamefont {S.}~\bibnamefont
  {{Yokoyama}}}, \bibinfo {author} {\bibfnamefont {T.}~\bibnamefont
  {{Ebihara}}}, \bibinfo {author} {\bibfnamefont {T.}~\bibnamefont
  {{Nakamura}}}, \bibinfo {author} {\bibfnamefont {R.~N.}\ \bibnamefont
  {{Alexander}}}, \bibinfo {author} {\bibfnamefont {M.}~\bibnamefont {{Endo}}},
  \bibinfo {author} {\bibfnamefont {J.}~\bibnamefont {{Yoshikawa}}}, \bibinfo
  {author} {\bibfnamefont {N.~C.}\ \bibnamefont {{Menicucci}}}, \bibinfo
  {author} {\bibfnamefont {H.}~\bibnamefont {{Yonezawa}}},\ and\ \bibinfo
  {author} {\bibfnamefont {A.}~\bibnamefont {{Furusawa}}},\ }\bibfield  {title}
  {\bibinfo {title} {One-hundred step measurement-based quantum computation
  multiplexed in the time domain with 25 {MHz} clock frequency},\ }\href@noop
  {} {\bibfield  {journal} {\bibinfo  {journal} {arXiv e-prints}\ ,\ \bibinfo
  {eid} {arXiv:2006.11537}} (\bibinfo {year} {2020})}\BibitemShut {NoStop}%
\bibitem [{\citenamefont {{Larsen}}\ \emph {et~al.}(2020)\citenamefont
  {{Larsen}}, \citenamefont {{Guo}}, \citenamefont {{Breum}}, \citenamefont
  {{Neergaard-Nielsen}},\ and\ \citenamefont
  {{Andersen}}}]{2020arXiv201014422L}%
  \BibitemOpen
  \bibfield  {author} {\bibinfo {author} {\bibfnamefont {M.~V.}\ \bibnamefont
  {{Larsen}}}, \bibinfo {author} {\bibfnamefont {X.}~\bibnamefont {{Guo}}},
  \bibinfo {author} {\bibfnamefont {C.~R.}\ \bibnamefont {{Breum}}}, \bibinfo
  {author} {\bibfnamefont {J.~S.}\ \bibnamefont {{Neergaard-Nielsen}}},\ and\
  \bibinfo {author} {\bibfnamefont {U.~L.}\ \bibnamefont {{Andersen}}},\
  }\bibfield  {title} {\bibinfo {title} {Deterministic multi-mode gates on a
  scalable photonic quantum computing platform},\ }\href@noop {} {\bibfield
  {journal} {\bibinfo  {journal} {arXiv e-prints}\ ,\ \bibinfo {eid}
  {arXiv:2010.14422}} (\bibinfo {year} {2020})}\BibitemShut {NoStop}%
\bibitem [{\citenamefont {Lloyd}\ and\ \citenamefont
  {Braunstein}(1999)}]{PhysRevLett.82.1784}%
  \BibitemOpen
  \bibfield  {author} {\bibinfo {author} {\bibfnamefont {S.}~\bibnamefont
  {Lloyd}}\ and\ \bibinfo {author} {\bibfnamefont {S.~L.}\ \bibnamefont
  {Braunstein}},\ }\bibfield  {title} {\bibinfo {title} {Quantum computation
  over continuous variables},\ }\href
  {https://doi.org/10.1103/PhysRevLett.82.1784} {\bibfield  {journal} {\bibinfo
   {journal} {Phys. Rev. Lett.}\ }\textbf {\bibinfo {volume} {82}},\ \bibinfo
  {pages} {1784} (\bibinfo {year} {1999})}\BibitemShut {NoStop}%
\bibitem [{\citenamefont {Bartlett}\ \emph {et~al.}(2002)\citenamefont
  {Bartlett}, \citenamefont {Sanders}, \citenamefont {Braunstein},\ and\
  \citenamefont {Nemoto}}]{PhysRevLett.88.097904}%
  \BibitemOpen
  \bibfield  {author} {\bibinfo {author} {\bibfnamefont {S.~D.}\ \bibnamefont
  {Bartlett}}, \bibinfo {author} {\bibfnamefont {B.~C.}\ \bibnamefont
  {Sanders}}, \bibinfo {author} {\bibfnamefont {S.~L.}\ \bibnamefont
  {Braunstein}},\ and\ \bibinfo {author} {\bibfnamefont {K.}~\bibnamefont
  {Nemoto}},\ }\bibfield  {title} {\bibinfo {title} {Efficient classical
  simulation of continuous variable quantum information processes},\ }\href
  {https://doi.org/10.1103/PhysRevLett.88.097904} {\bibfield  {journal}
  {\bibinfo  {journal} {Phys. Rev. Lett.}\ }\textbf {\bibinfo {volume} {88}},\
  \bibinfo {pages} {097904} (\bibinfo {year} {2002})}\BibitemShut {NoStop}%
\bibitem [{\citenamefont {Gottesman}\ and\ \citenamefont
  {Chuang}(1999)}]{Gottesman1999}%
  \BibitemOpen
  \bibfield  {author} {\bibinfo {author} {\bibfnamefont {D.}~\bibnamefont
  {Gottesman}}\ and\ \bibinfo {author} {\bibfnamefont {I.~L.}\ \bibnamefont
  {Chuang}},\ }\bibfield  {title} {\bibinfo {title} {Demonstrating the
  viability of universal quantum computation using teleportation and
  single-qubit operations},\ }\href {https://doi.org/10.1038/46503} {\bibfield
  {journal} {\bibinfo  {journal} {Nature}\ }\textbf {\bibinfo {volume} {402}},\
  \bibinfo {pages} {390} (\bibinfo {year} {1999})}\BibitemShut {NoStop}%
\bibitem [{\citenamefont {Gottesman}\ \emph {et~al.}(2001)\citenamefont
  {Gottesman}, \citenamefont {Kitaev},\ and\ \citenamefont
  {Preskill}}]{PhysRevA.64.012310}%
  \BibitemOpen
  \bibfield  {author} {\bibinfo {author} {\bibfnamefont {D.}~\bibnamefont
  {Gottesman}}, \bibinfo {author} {\bibfnamefont {A.}~\bibnamefont {Kitaev}},\
  and\ \bibinfo {author} {\bibfnamefont {J.}~\bibnamefont {Preskill}},\
  }\bibfield  {title} {\bibinfo {title} {Encoding a qubit in an oscillator},\
  }\href {https://doi.org/10.1103/PhysRevA.64.012310} {\bibfield  {journal}
  {\bibinfo  {journal} {Phys. Rev. A}\ }\textbf {\bibinfo {volume} {64}},\
  \bibinfo {pages} {012310} (\bibinfo {year} {2001})}\BibitemShut {NoStop}%
\bibitem [{\citenamefont {Cochrane}\ \emph {et~al.}(1999)\citenamefont
  {Cochrane}, \citenamefont {Milburn},\ and\ \citenamefont
  {Munro}}]{PhysRevA.59.2631}%
  \BibitemOpen
  \bibfield  {author} {\bibinfo {author} {\bibfnamefont {P.~T.}\ \bibnamefont
  {Cochrane}}, \bibinfo {author} {\bibfnamefont {G.~J.}\ \bibnamefont
  {Milburn}},\ and\ \bibinfo {author} {\bibfnamefont {W.~J.}\ \bibnamefont
  {Munro}},\ }\bibfield  {title} {\bibinfo {title} {Macroscopically distinct
  quantum-superposition states as a bosonic code for amplitude damping},\
  }\href {https://doi.org/10.1103/PhysRevA.59.2631} {\bibfield  {journal}
  {\bibinfo  {journal} {Phys. Rev. A}\ }\textbf {\bibinfo {volume} {59}},\
  \bibinfo {pages} {2631} (\bibinfo {year} {1999})}\BibitemShut {NoStop}%
\bibitem [{\citenamefont {Ralph}\ \emph {et~al.}(2003)\citenamefont {Ralph},
  \citenamefont {Gilchrist}, \citenamefont {Milburn}, \citenamefont {Munro},\
  and\ \citenamefont {Glancy}}]{PhysRevA.68.042319}%
  \BibitemOpen
  \bibfield  {author} {\bibinfo {author} {\bibfnamefont {T.~C.}\ \bibnamefont
  {Ralph}}, \bibinfo {author} {\bibfnamefont {A.}~\bibnamefont {Gilchrist}},
  \bibinfo {author} {\bibfnamefont {G.~J.}\ \bibnamefont {Milburn}}, \bibinfo
  {author} {\bibfnamefont {W.~J.}\ \bibnamefont {Munro}},\ and\ \bibinfo
  {author} {\bibfnamefont {S.}~\bibnamefont {Glancy}},\ }\bibfield  {title}
  {\bibinfo {title} {Quantum computation with optical coherent states},\ }\href
  {https://doi.org/10.1103/PhysRevA.68.042319} {\bibfield  {journal} {\bibinfo
  {journal} {Phys. Rev. A}\ }\textbf {\bibinfo {volume} {68}},\ \bibinfo
  {pages} {042319} (\bibinfo {year} {2003})}\BibitemShut {NoStop}%
\bibitem [{\citenamefont {Lund}\ \emph {et~al.}(2008)\citenamefont {Lund},
  \citenamefont {Ralph},\ and\ \citenamefont
  {Haselgrove}}]{PhysRevLett.100.030503}%
  \BibitemOpen
  \bibfield  {author} {\bibinfo {author} {\bibfnamefont {A.~P.}\ \bibnamefont
  {Lund}}, \bibinfo {author} {\bibfnamefont {T.~C.}\ \bibnamefont {Ralph}},\
  and\ \bibinfo {author} {\bibfnamefont {H.~L.}\ \bibnamefont {Haselgrove}},\
  }\bibfield  {title} {\bibinfo {title} {Fault-tolerant linear optical quantum
  computing with small-amplitude coherent states},\ }\href
  {https://doi.org/10.1103/PhysRevLett.100.030503} {\bibfield  {journal}
  {\bibinfo  {journal} {Phys. Rev. Lett.}\ }\textbf {\bibinfo {volume} {100}},\
  \bibinfo {pages} {030503} (\bibinfo {year} {2008})}\BibitemShut {NoStop}%
\bibitem [{\citenamefont {Neergaard-Nielsen}\ \emph {et~al.}(2010)\citenamefont
  {Neergaard-Nielsen}, \citenamefont {Takeuchi}, \citenamefont {Wakui},
  \citenamefont {Takahashi}, \citenamefont {Hayasaka}, \citenamefont
  {Takeoka},\ and\ \citenamefont {Sasaki}}]{PhysRevLett.105.053602}%
  \BibitemOpen
  \bibfield  {author} {\bibinfo {author} {\bibfnamefont {J.~S.}\ \bibnamefont
  {Neergaard-Nielsen}}, \bibinfo {author} {\bibfnamefont {M.}~\bibnamefont
  {Takeuchi}}, \bibinfo {author} {\bibfnamefont {K.}~\bibnamefont {Wakui}},
  \bibinfo {author} {\bibfnamefont {H.}~\bibnamefont {Takahashi}}, \bibinfo
  {author} {\bibfnamefont {K.}~\bibnamefont {Hayasaka}}, \bibinfo {author}
  {\bibfnamefont {M.}~\bibnamefont {Takeoka}},\ and\ \bibinfo {author}
  {\bibfnamefont {M.}~\bibnamefont {Sasaki}},\ }\bibfield  {title} {\bibinfo
  {title} {Optical continuous-variable qubit},\ }\href
  {https://doi.org/10.1103/PhysRevLett.105.053602} {\bibfield  {journal}
  {\bibinfo  {journal} {Phys. Rev. Lett.}\ }\textbf {\bibinfo {volume} {105}},\
  \bibinfo {pages} {053602} (\bibinfo {year} {2010})}\BibitemShut {NoStop}%
\bibitem [{\citenamefont {Karimipour}\ \emph {et~al.}(2002)\citenamefont
  {Karimipour}, \citenamefont {Bahraminasab},\ and\ \citenamefont
  {Bagherinezhad}}]{PhysRevA.65.042320}%
  \BibitemOpen
  \bibfield  {author} {\bibinfo {author} {\bibfnamefont {V.}~\bibnamefont
  {Karimipour}}, \bibinfo {author} {\bibfnamefont {A.}~\bibnamefont
  {Bahraminasab}},\ and\ \bibinfo {author} {\bibfnamefont {S.}~\bibnamefont
  {Bagherinezhad}},\ }\bibfield  {title} {\bibinfo {title} {Entanglement
  swapping of generalized cat states and secret sharing},\ }\href
  {https://doi.org/10.1103/PhysRevA.65.042320} {\bibfield  {journal} {\bibinfo
  {journal} {Phys. Rev. A}\ }\textbf {\bibinfo {volume} {65}},\ \bibinfo
  {pages} {042320} (\bibinfo {year} {2002})}\BibitemShut {NoStop}%
\bibitem [{\citenamefont {Sangouard}\ \emph {et~al.}(2010)\citenamefont
  {Sangouard}, \citenamefont {Simon}, \citenamefont {Gisin}, \citenamefont
  {Laurat}, \citenamefont {Tualle-Brouri},\ and\ \citenamefont
  {Grangier}}]{Sangouard:10}%
  \BibitemOpen
  \bibfield  {author} {\bibinfo {author} {\bibfnamefont {N.}~\bibnamefont
  {Sangouard}}, \bibinfo {author} {\bibfnamefont {C.}~\bibnamefont {Simon}},
  \bibinfo {author} {\bibfnamefont {N.}~\bibnamefont {Gisin}}, \bibinfo
  {author} {\bibfnamefont {J.}~\bibnamefont {Laurat}}, \bibinfo {author}
  {\bibfnamefont {R.}~\bibnamefont {Tualle-Brouri}},\ and\ \bibinfo {author}
  {\bibfnamefont {P.}~\bibnamefont {Grangier}},\ }\bibfield  {title} {\bibinfo
  {title} {Quantum repeaters with entangled coherent states},\ }\href
  {https://doi.org/10.1364/JOSAB.27.00A137} {\bibfield  {journal} {\bibinfo
  {journal} {J. Opt. Soc. Am. B}\ }\textbf {\bibinfo {volume} {27}},\ \bibinfo
  {pages} {A137} (\bibinfo {year} {2010})}\BibitemShut {NoStop}%
\bibitem [{\citenamefont {Vasconcelos}\ \emph {et~al.}(2010)\citenamefont
  {Vasconcelos}, \citenamefont {Sanz},\ and\ \citenamefont
  {Glancy}}]{Vasconcelos:10}%
  \BibitemOpen
  \bibfield  {author} {\bibinfo {author} {\bibfnamefont {H.~M.}\ \bibnamefont
  {Vasconcelos}}, \bibinfo {author} {\bibfnamefont {L.}~\bibnamefont {Sanz}},\
  and\ \bibinfo {author} {\bibfnamefont {S.}~\bibnamefont {Glancy}},\
  }\bibfield  {title} {\bibinfo {title} {All-optical generation of states for
  ``encoding a qubit in an oscillator''},\ }\href
  {https://doi.org/10.1364/OL.35.003261} {\bibfield  {journal} {\bibinfo
  {journal} {Opt. Lett.}\ }\textbf {\bibinfo {volume} {35}},\ \bibinfo {pages}
  {3261} (\bibinfo {year} {2010})}\BibitemShut {NoStop}%
\bibitem [{\citenamefont {Hastrup}\ \emph {et~al.}(2020)\citenamefont
  {Hastrup}, \citenamefont {Neergaard-Nielsen},\ and\ \citenamefont
  {Andersen}}]{Hastrup:20}%
  \BibitemOpen
  \bibfield  {author} {\bibinfo {author} {\bibfnamefont {J.}~\bibnamefont
  {Hastrup}}, \bibinfo {author} {\bibfnamefont {J.~S.}\ \bibnamefont
  {Neergaard-Nielsen}},\ and\ \bibinfo {author} {\bibfnamefont {U.~L.}\
  \bibnamefont {Andersen}},\ }\bibfield  {title} {\bibinfo {title}
  {Deterministic generation of a four-component optical cat state},\ }\href
  {https://doi.org/10.1364/OL.383194} {\bibfield  {journal} {\bibinfo
  {journal} {Opt. Lett.}\ }\textbf {\bibinfo {volume} {45}},\ \bibinfo {pages}
  {640} (\bibinfo {year} {2020})}\BibitemShut {NoStop}%
\bibitem [{\citenamefont {Weigand}\ and\ \citenamefont
  {Terhal}(2018)}]{PhysRevA.97.022341}%
  \BibitemOpen
  \bibfield  {author} {\bibinfo {author} {\bibfnamefont {D.~J.}\ \bibnamefont
  {Weigand}}\ and\ \bibinfo {author} {\bibfnamefont {B.~M.}\ \bibnamefont
  {Terhal}},\ }\bibfield  {title} {\bibinfo {title} {Generating grid states
  from schr\"odinger-cat states without postselection},\ }\href
  {https://doi.org/10.1103/PhysRevA.97.022341} {\bibfield  {journal} {\bibinfo
  {journal} {Phys. Rev. A}\ }\textbf {\bibinfo {volume} {97}},\ \bibinfo
  {pages} {022341} (\bibinfo {year} {2018})}\BibitemShut {NoStop}%
\bibitem [{\citenamefont {Menicucci}(2014)}]{PhysRevLett.112.120504}%
  \BibitemOpen
  \bibfield  {author} {\bibinfo {author} {\bibfnamefont {N.~C.}\ \bibnamefont
  {Menicucci}},\ }\bibfield  {title} {\bibinfo {title} {Fault-tolerant
  measurement-based quantum computing with continuous-variable cluster
  states},\ }\href {https://doi.org/10.1103/PhysRevLett.112.120504} {\bibfield
  {journal} {\bibinfo  {journal} {Phys. Rev. Lett.}\ }\textbf {\bibinfo
  {volume} {112}},\ \bibinfo {pages} {120504} (\bibinfo {year}
  {2014})}\BibitemShut {NoStop}%
\bibitem [{\citenamefont {Fukui}\ \emph {et~al.}(2018)\citenamefont {Fukui},
  \citenamefont {Tomita}, \citenamefont {Okamoto},\ and\ \citenamefont
  {Fujii}}]{PhysRevX.8.021054}%
  \BibitemOpen
  \bibfield  {author} {\bibinfo {author} {\bibfnamefont {K.}~\bibnamefont
  {Fukui}}, \bibinfo {author} {\bibfnamefont {A.}~\bibnamefont {Tomita}},
  \bibinfo {author} {\bibfnamefont {A.}~\bibnamefont {Okamoto}},\ and\ \bibinfo
  {author} {\bibfnamefont {K.}~\bibnamefont {Fujii}},\ }\bibfield  {title}
  {\bibinfo {title} {High-threshold fault-tolerant quantum computation with
  analog quantum error correction},\ }\href
  {https://doi.org/10.1103/PhysRevX.8.021054} {\bibfield  {journal} {\bibinfo
  {journal} {Phys. Rev. X}\ }\textbf {\bibinfo {volume} {8}},\ \bibinfo {pages}
  {021054} (\bibinfo {year} {2018})}\BibitemShut {NoStop}%
\bibitem [{\citenamefont {Walshe}\ \emph {et~al.}(2019)\citenamefont {Walshe},
  \citenamefont {Mensen}, \citenamefont {Baragiola},\ and\ \citenamefont
  {Menicucci}}]{PhysRevA.100.010301}%
  \BibitemOpen
  \bibfield  {author} {\bibinfo {author} {\bibfnamefont {B.~W.}\ \bibnamefont
  {Walshe}}, \bibinfo {author} {\bibfnamefont {L.~J.}\ \bibnamefont {Mensen}},
  \bibinfo {author} {\bibfnamefont {B.~Q.}\ \bibnamefont {Baragiola}},\ and\
  \bibinfo {author} {\bibfnamefont {N.~C.}\ \bibnamefont {Menicucci}},\
  }\bibfield  {title} {\bibinfo {title} {Robust fault tolerance for
  continuous-variable cluster states with excess antisqueezing},\ }\href
  {https://doi.org/10.1103/PhysRevA.100.010301} {\bibfield  {journal} {\bibinfo
   {journal} {Phys. Rev. A}\ }\textbf {\bibinfo {volume} {100}},\ \bibinfo
  {pages} {010301} (\bibinfo {year} {2019})}\BibitemShut {NoStop}%
\bibitem [{\citenamefont {Baragiola}\ \emph {et~al.}(2019)\citenamefont
  {Baragiola}, \citenamefont {Pantaleoni}, \citenamefont {Alexander},
  \citenamefont {Karanjai},\ and\ \citenamefont
  {Menicucci}}]{PhysRevLett.123.200502}%
  \BibitemOpen
  \bibfield  {author} {\bibinfo {author} {\bibfnamefont {B.~Q.}\ \bibnamefont
  {Baragiola}}, \bibinfo {author} {\bibfnamefont {G.}~\bibnamefont
  {Pantaleoni}}, \bibinfo {author} {\bibfnamefont {R.~N.}\ \bibnamefont
  {Alexander}}, \bibinfo {author} {\bibfnamefont {A.}~\bibnamefont
  {Karanjai}},\ and\ \bibinfo {author} {\bibfnamefont {N.~C.}\ \bibnamefont
  {Menicucci}},\ }\bibfield  {title} {\bibinfo {title} {All-gaussian
  universality and fault tolerance with the gottesman-kitaev-preskill code},\
  }\href {https://doi.org/10.1103/PhysRevLett.123.200502} {\bibfield  {journal}
  {\bibinfo  {journal} {Phys. Rev. Lett.}\ }\textbf {\bibinfo {volume} {123}},\
  \bibinfo {pages} {200502} (\bibinfo {year} {2019})}\BibitemShut {NoStop}%
\bibitem [{\citenamefont {Yamasaki}\ \emph {et~al.}(2020)\citenamefont
  {Yamasaki}, \citenamefont {Matsuura},\ and\ \citenamefont
  {Koashi}}]{PhysRevResearch.2.023270}%
  \BibitemOpen
  \bibfield  {author} {\bibinfo {author} {\bibfnamefont {H.}~\bibnamefont
  {Yamasaki}}, \bibinfo {author} {\bibfnamefont {T.}~\bibnamefont {Matsuura}},\
  and\ \bibinfo {author} {\bibfnamefont {M.}~\bibnamefont {Koashi}},\
  }\bibfield  {title} {\bibinfo {title} {Cost-reduced all-gaussian universality
  with the gottesman-kitaev-preskill code: Resource-theoretic approach to cost
  analysis},\ }\href {https://doi.org/10.1103/PhysRevResearch.2.023270}
  {\bibfield  {journal} {\bibinfo  {journal} {Phys. Rev. Research}\ }\textbf
  {\bibinfo {volume} {2}},\ \bibinfo {pages} {023270} (\bibinfo {year}
  {2020})}\BibitemShut {NoStop}%
\bibitem [{\citenamefont {Tzitrin}\ \emph {et~al.}(2020)\citenamefont
  {Tzitrin}, \citenamefont {Bourassa}, \citenamefont {Menicucci},\ and\
  \citenamefont {Sabapathy}}]{PhysRevA.101.032315}%
  \BibitemOpen
  \bibfield  {author} {\bibinfo {author} {\bibfnamefont {I.}~\bibnamefont
  {Tzitrin}}, \bibinfo {author} {\bibfnamefont {J.~E.}\ \bibnamefont
  {Bourassa}}, \bibinfo {author} {\bibfnamefont {N.~C.}\ \bibnamefont
  {Menicucci}},\ and\ \bibinfo {author} {\bibfnamefont {K.~K.}\ \bibnamefont
  {Sabapathy}},\ }\bibfield  {title} {\bibinfo {title} {Progress towards
  practical qubit computation using approximate gottesman-kitaev-preskill
  codes},\ }\href {https://doi.org/10.1103/PhysRevA.101.032315} {\bibfield
  {journal} {\bibinfo  {journal} {Phys. Rev. A}\ }\textbf {\bibinfo {volume}
  {101}},\ \bibinfo {pages} {032315} (\bibinfo {year} {2020})}\BibitemShut
  {NoStop}%
\bibitem [{\citenamefont {Lvovsky}\ and\ \citenamefont
  {Mlynek}(2002)}]{PhysRevLett.88.250401}%
  \BibitemOpen
  \bibfield  {author} {\bibinfo {author} {\bibfnamefont {A.~I.}\ \bibnamefont
  {Lvovsky}}\ and\ \bibinfo {author} {\bibfnamefont {J.}~\bibnamefont
  {Mlynek}},\ }\bibfield  {title} {\bibinfo {title} {Quantum-optical catalysis:
  Generating nonclassical states of light by means of linear optics},\ }\href
  {https://doi.org/10.1103/PhysRevLett.88.250401} {\bibfield  {journal}
  {\bibinfo  {journal} {Phys. Rev. Lett.}\ }\textbf {\bibinfo {volume} {88}},\
  \bibinfo {pages} {250401} (\bibinfo {year} {2002})}\BibitemShut {NoStop}%
\bibitem [{\citenamefont {Resch}\ \emph {et~al.}(2002)\citenamefont {Resch},
  \citenamefont {Lundeen},\ and\ \citenamefont
  {Steinberg}}]{PhysRevLett.88.113601}%
  \BibitemOpen
  \bibfield  {author} {\bibinfo {author} {\bibfnamefont {K.~J.}\ \bibnamefont
  {Resch}}, \bibinfo {author} {\bibfnamefont {J.~S.}\ \bibnamefont {Lundeen}},\
  and\ \bibinfo {author} {\bibfnamefont {A.~M.}\ \bibnamefont {Steinberg}},\
  }\bibfield  {title} {\bibinfo {title} {Quantum state preparation and
  conditional coherence},\ }\href
  {https://doi.org/10.1103/PhysRevLett.88.113601} {\bibfield  {journal}
  {\bibinfo  {journal} {Phys. Rev. Lett.}\ }\textbf {\bibinfo {volume} {88}},\
  \bibinfo {pages} {113601} (\bibinfo {year} {2002})}\BibitemShut {NoStop}%
\bibitem [{\citenamefont {Hashimoto}\ \emph {et~al.}(2019)\citenamefont
  {Hashimoto}, \citenamefont {Toyama}, \citenamefont {Yoshikawa}, \citenamefont
  {Makino}, \citenamefont {Okamoto}, \citenamefont {Sakakibara}, \citenamefont
  {Takeda}, \citenamefont {van Loock},\ and\ \citenamefont
  {Furusawa}}]{PhysRevLett.123.113603}%
  \BibitemOpen
  \bibfield  {author} {\bibinfo {author} {\bibfnamefont {Y.}~\bibnamefont
  {Hashimoto}}, \bibinfo {author} {\bibfnamefont {T.}~\bibnamefont {Toyama}},
  \bibinfo {author} {\bibfnamefont {J.}~\bibnamefont {Yoshikawa}}, \bibinfo
  {author} {\bibfnamefont {K.}~\bibnamefont {Makino}}, \bibinfo {author}
  {\bibfnamefont {F.}~\bibnamefont {Okamoto}}, \bibinfo {author} {\bibfnamefont
  {R.}~\bibnamefont {Sakakibara}}, \bibinfo {author} {\bibfnamefont
  {S.}~\bibnamefont {Takeda}}, \bibinfo {author} {\bibfnamefont
  {P.}~\bibnamefont {van Loock}},\ and\ \bibinfo {author} {\bibfnamefont
  {A.}~\bibnamefont {Furusawa}},\ }\bibfield  {title} {\bibinfo {title}
  {All-optical storage of phase-sensitive quantum states of light},\ }\href
  {https://doi.org/10.1103/PhysRevLett.123.113603} {\bibfield  {journal}
  {\bibinfo  {journal} {Phys. Rev. Lett.}\ }\textbf {\bibinfo {volume} {123}},\
  \bibinfo {pages} {113603} (\bibinfo {year} {2019})}\BibitemShut {NoStop}%
\bibitem [{\citenamefont {Bimbard}\ \emph {et~al.}(2010)\citenamefont
  {Bimbard}, \citenamefont {Jain}, \citenamefont {MacRae},\ and\ \citenamefont
  {Lvovsky}}]{Bimbard2010}%
  \BibitemOpen
  \bibfield  {author} {\bibinfo {author} {\bibfnamefont {E.}~\bibnamefont
  {Bimbard}}, \bibinfo {author} {\bibfnamefont {N.}~\bibnamefont {Jain}},
  \bibinfo {author} {\bibfnamefont {A.}~\bibnamefont {MacRae}},\ and\ \bibinfo
  {author} {\bibfnamefont {A.~I.}\ \bibnamefont {Lvovsky}},\ }\bibfield
  {title} {\bibinfo {title} {Quantum-optical state engineering up to the
  two-photon level},\ }\href@noop {} {\bibfield  {journal} {\bibinfo  {journal}
  {Nature Photonics}\ }\textbf {\bibinfo {volume} {4}},\ \bibinfo {pages} {243}
  (\bibinfo {year} {2010})}\BibitemShut {NoStop}%
\bibitem [{\citenamefont {Bartley}\ \emph {et~al.}(2012)\citenamefont
  {Bartley}, \citenamefont {Donati}, \citenamefont {Spring}, \citenamefont
  {Jin}, \citenamefont {Barbieri}, \citenamefont {Datta}, \citenamefont
  {Smith},\ and\ \citenamefont {Walmsley}}]{PhysRevA.86.043820}%
  \BibitemOpen
  \bibfield  {author} {\bibinfo {author} {\bibfnamefont {T.~J.}\ \bibnamefont
  {Bartley}}, \bibinfo {author} {\bibfnamefont {G.}~\bibnamefont {Donati}},
  \bibinfo {author} {\bibfnamefont {J.~B.}\ \bibnamefont {Spring}}, \bibinfo
  {author} {\bibfnamefont {X.-M.}\ \bibnamefont {Jin}}, \bibinfo {author}
  {\bibfnamefont {M.}~\bibnamefont {Barbieri}}, \bibinfo {author}
  {\bibfnamefont {A.}~\bibnamefont {Datta}}, \bibinfo {author} {\bibfnamefont
  {B.~J.}\ \bibnamefont {Smith}},\ and\ \bibinfo {author} {\bibfnamefont
  {I.~A.}\ \bibnamefont {Walmsley}},\ }\bibfield  {title} {\bibinfo {title}
  {Multiphoton state engineering by heralded interference between single
  photons and coherent states},\ }\href
  {https://doi.org/10.1103/PhysRevA.86.043820} {\bibfield  {journal} {\bibinfo
  {journal} {Phys. Rev. A}\ }\textbf {\bibinfo {volume} {86}},\ \bibinfo
  {pages} {043820} (\bibinfo {year} {2012})}\BibitemShut {NoStop}%
\bibitem [{\citenamefont {Yukawa}\ \emph {et~al.}(2013)\citenamefont {Yukawa},
  \citenamefont {Miyata}, \citenamefont {Mizuta}, \citenamefont {Yonezawa},
  \citenamefont {Marek}, \citenamefont {Filip},\ and\ \citenamefont
  {Furusawa}}]{Yukawa:13}%
  \BibitemOpen
  \bibfield  {author} {\bibinfo {author} {\bibfnamefont {M.}~\bibnamefont
  {Yukawa}}, \bibinfo {author} {\bibfnamefont {K.}~\bibnamefont {Miyata}},
  \bibinfo {author} {\bibfnamefont {T.}~\bibnamefont {Mizuta}}, \bibinfo
  {author} {\bibfnamefont {H.}~\bibnamefont {Yonezawa}}, \bibinfo {author}
  {\bibfnamefont {P.}~\bibnamefont {Marek}}, \bibinfo {author} {\bibfnamefont
  {R.}~\bibnamefont {Filip}},\ and\ \bibinfo {author} {\bibfnamefont
  {A.}~\bibnamefont {Furusawa}},\ }\bibfield  {title} {\bibinfo {title}
  {Generating superposition of up-to three photons for continuous variable
  quantum information processing},\ }\href
  {https://doi.org/10.1364/OE.21.005529} {\bibfield  {journal} {\bibinfo
  {journal} {Opt. Express}\ }\textbf {\bibinfo {volume} {21}},\ \bibinfo
  {pages} {5529} (\bibinfo {year} {2013})}\BibitemShut {NoStop}%
\bibitem [{\citenamefont {Dakna}\ \emph {et~al.}(1997)\citenamefont {Dakna},
  \citenamefont {Anhut}, \citenamefont {Opatrn\'y}, \citenamefont {Kn\"oll},\
  and\ \citenamefont {Welsch}}]{PhysRevA.55.3184}%
  \BibitemOpen
  \bibfield  {author} {\bibinfo {author} {\bibfnamefont {M.}~\bibnamefont
  {Dakna}}, \bibinfo {author} {\bibfnamefont {T.}~\bibnamefont {Anhut}},
  \bibinfo {author} {\bibfnamefont {T.}~\bibnamefont {Opatrn\'y}}, \bibinfo
  {author} {\bibfnamefont {L.}~\bibnamefont {Kn\"oll}},\ and\ \bibinfo {author}
  {\bibfnamefont {D.-G.}\ \bibnamefont {Welsch}},\ }\bibfield  {title}
  {\bibinfo {title} {Generating schr\"odinger-cat-like states by means of
  conditional measurements on a beam splitter},\ }\href
  {https://doi.org/10.1103/PhysRevA.55.3184} {\bibfield  {journal} {\bibinfo
  {journal} {Phys. Rev. A}\ }\textbf {\bibinfo {volume} {55}},\ \bibinfo
  {pages} {3184} (\bibinfo {year} {1997})}\BibitemShut {NoStop}%
\bibitem [{\citenamefont {Huang}\ \emph {et~al.}(2015)\citenamefont {Huang},
  \citenamefont {Le~Jeannic}, \citenamefont {Ruaudel}, \citenamefont {Verma},
  \citenamefont {Shaw}, \citenamefont {Marsili}, \citenamefont {Nam},
  \citenamefont {Wu}, \citenamefont {Zeng}, \citenamefont {Jeong},
  \citenamefont {Filip}, \citenamefont {Morin},\ and\ \citenamefont
  {Laurat}}]{PhysRevLett.115.023602}%
  \BibitemOpen
  \bibfield  {author} {\bibinfo {author} {\bibfnamefont {K.}~\bibnamefont
  {Huang}}, \bibinfo {author} {\bibfnamefont {H.}~\bibnamefont {Le~Jeannic}},
  \bibinfo {author} {\bibfnamefont {J.}~\bibnamefont {Ruaudel}}, \bibinfo
  {author} {\bibfnamefont {V.~B.}\ \bibnamefont {Verma}}, \bibinfo {author}
  {\bibfnamefont {M.~D.}\ \bibnamefont {Shaw}}, \bibinfo {author}
  {\bibfnamefont {F.}~\bibnamefont {Marsili}}, \bibinfo {author} {\bibfnamefont
  {S.~W.}\ \bibnamefont {Nam}}, \bibinfo {author} {\bibfnamefont
  {E.}~\bibnamefont {Wu}}, \bibinfo {author} {\bibfnamefont {H.}~\bibnamefont
  {Zeng}}, \bibinfo {author} {\bibfnamefont {Y.-C.}\ \bibnamefont {Jeong}},
  \bibinfo {author} {\bibfnamefont {R.}~\bibnamefont {Filip}}, \bibinfo
  {author} {\bibfnamefont {O.}~\bibnamefont {Morin}},\ and\ \bibinfo {author}
  {\bibfnamefont {J.}~\bibnamefont {Laurat}},\ }\bibfield  {title} {\bibinfo
  {title} {Optical synthesis of large-amplitude squeezed coherent-state
  superpositions with minimal resources},\ }\href
  {https://doi.org/10.1103/PhysRevLett.115.023602} {\bibfield  {journal}
  {\bibinfo  {journal} {Phys. Rev. Lett.}\ }\textbf {\bibinfo {volume} {115}},\
  \bibinfo {pages} {023602} (\bibinfo {year} {2015})}\BibitemShut {NoStop}%
\bibitem [{\citenamefont {Gerrits}\ \emph {et~al.}(2010)\citenamefont
  {Gerrits}, \citenamefont {Glancy}, \citenamefont {Clement}, \citenamefont
  {Calkins}, \citenamefont {Lita}, \citenamefont {Miller}, \citenamefont
  {Migdall}, \citenamefont {Nam}, \citenamefont {Mirin},\ and\ \citenamefont
  {Knill}}]{PhysRevA.82.031802}%
  \BibitemOpen
  \bibfield  {author} {\bibinfo {author} {\bibfnamefont {T.}~\bibnamefont
  {Gerrits}}, \bibinfo {author} {\bibfnamefont {S.}~\bibnamefont {Glancy}},
  \bibinfo {author} {\bibfnamefont {T.~S.}\ \bibnamefont {Clement}}, \bibinfo
  {author} {\bibfnamefont {B.}~\bibnamefont {Calkins}}, \bibinfo {author}
  {\bibfnamefont {A.~E.}\ \bibnamefont {Lita}}, \bibinfo {author}
  {\bibfnamefont {A.~J.}\ \bibnamefont {Miller}}, \bibinfo {author}
  {\bibfnamefont {A.~L.}\ \bibnamefont {Migdall}}, \bibinfo {author}
  {\bibfnamefont {S.~W.}\ \bibnamefont {Nam}}, \bibinfo {author} {\bibfnamefont
  {R.~P.}\ \bibnamefont {Mirin}},\ and\ \bibinfo {author} {\bibfnamefont
  {E.}~\bibnamefont {Knill}},\ }\bibfield  {title} {\bibinfo {title}
  {Generation of optical coherent-state superpositions by number-resolved
  photon subtraction from the squeezed vacuum},\ }\href
  {https://doi.org/10.1103/PhysRevA.82.031802} {\bibfield  {journal} {\bibinfo
  {journal} {Phys. Rev. A}\ }\textbf {\bibinfo {volume} {82}},\ \bibinfo
  {pages} {031802} (\bibinfo {year} {2010})}\BibitemShut {NoStop}%
\bibitem [{\citenamefont {Ourjoumtsev}\ \emph {et~al.}(2006)\citenamefont
  {Ourjoumtsev}, \citenamefont {Tualle-Brouri}, \citenamefont {Laurat},\ and\
  \citenamefont {Grangier}}]{Ourjoumtsev83}%
  \BibitemOpen
  \bibfield  {author} {\bibinfo {author} {\bibfnamefont {A.}~\bibnamefont
  {Ourjoumtsev}}, \bibinfo {author} {\bibfnamefont {R.}~\bibnamefont
  {Tualle-Brouri}}, \bibinfo {author} {\bibfnamefont {J.}~\bibnamefont
  {Laurat}},\ and\ \bibinfo {author} {\bibfnamefont {P.}~\bibnamefont
  {Grangier}},\ }\bibfield  {title} {\bibinfo {title} {Generating optical
  schr{\"o}dinger kittens for quantum information processing},\ }\href
  {https://doi.org/10.1126/science.1122858} {\bibfield  {journal} {\bibinfo
  {journal} {Science}\ }\textbf {\bibinfo {volume} {312}},\ \bibinfo {pages}
  {83} (\bibinfo {year} {2006})}\BibitemShut {NoStop}%
\bibitem [{\citenamefont {Neergaard-Nielsen}\ \emph {et~al.}(2006)\citenamefont
  {Neergaard-Nielsen}, \citenamefont {Nielsen}, \citenamefont {Hettich},
  \citenamefont {M\o{}lmer},\ and\ \citenamefont
  {Polzik}}]{PhysRevLett.97.083604}%
  \BibitemOpen
  \bibfield  {author} {\bibinfo {author} {\bibfnamefont {J.~S.}\ \bibnamefont
  {Neergaard-Nielsen}}, \bibinfo {author} {\bibfnamefont {B.~M.}\ \bibnamefont
  {Nielsen}}, \bibinfo {author} {\bibfnamefont {C.}~\bibnamefont {Hettich}},
  \bibinfo {author} {\bibfnamefont {K.}~\bibnamefont {M\o{}lmer}},\ and\
  \bibinfo {author} {\bibfnamefont {E.~S.}\ \bibnamefont {Polzik}},\ }\bibfield
   {title} {\bibinfo {title} {Generation of a superposition of odd photon
  number states for quantum information networks},\ }\href
  {https://doi.org/10.1103/PhysRevLett.97.083604} {\bibfield  {journal}
  {\bibinfo  {journal} {Phys. Rev. Lett.}\ }\textbf {\bibinfo {volume} {97}},\
  \bibinfo {pages} {083604} (\bibinfo {year} {2006})}\BibitemShut {NoStop}%
\bibitem [{\citenamefont {Wakui}\ \emph {et~al.}(2007)\citenamefont {Wakui},
  \citenamefont {Takahashi}, \citenamefont {Furusawa},\ and\ \citenamefont
  {Sasaki}}]{Wakui:07}%
  \BibitemOpen
  \bibfield  {author} {\bibinfo {author} {\bibfnamefont {K.}~\bibnamefont
  {Wakui}}, \bibinfo {author} {\bibfnamefont {H.}~\bibnamefont {Takahashi}},
  \bibinfo {author} {\bibfnamefont {A.}~\bibnamefont {Furusawa}},\ and\
  \bibinfo {author} {\bibfnamefont {M.}~\bibnamefont {Sasaki}},\ }\bibfield
  {title} {\bibinfo {title} {Photon subtracted squeezed states generated with
  periodically poled ktiopo4},\ }\href {https://doi.org/10.1364/OE.15.003568}
  {\bibfield  {journal} {\bibinfo  {journal} {Opt. Express}\ }\textbf {\bibinfo
  {volume} {15}},\ \bibinfo {pages} {3568} (\bibinfo {year}
  {2007})}\BibitemShut {NoStop}%
\bibitem [{\citenamefont {Park}\ \emph {et~al.}(2014)\citenamefont {Park},
  \citenamefont {Marek},\ and\ \citenamefont {Filip}}]{PhysRevA.90.013804}%
  \BibitemOpen
  \bibfield  {author} {\bibinfo {author} {\bibfnamefont {K.}~\bibnamefont
  {Park}}, \bibinfo {author} {\bibfnamefont {P.}~\bibnamefont {Marek}},\ and\
  \bibinfo {author} {\bibfnamefont {R.}~\bibnamefont {Filip}},\ }\bibfield
  {title} {\bibinfo {title} {Nonlinear potential of a quantum oscillator
  induced by single photons},\ }\href
  {https://doi.org/10.1103/PhysRevA.90.013804} {\bibfield  {journal} {\bibinfo
  {journal} {Phys. Rev. A}\ }\textbf {\bibinfo {volume} {90}},\ \bibinfo
  {pages} {013804} (\bibinfo {year} {2014})}\BibitemShut {NoStop}%
\bibitem [{\citenamefont {Ourjoumtsev}\ \emph {et~al.}(2007)\citenamefont
  {Ourjoumtsev}, \citenamefont {Jeong}, \citenamefont {Tualle-Brouri},\ and\
  \citenamefont {Grangier}}]{Ourjoumtsev2007}%
  \BibitemOpen
  \bibfield  {author} {\bibinfo {author} {\bibfnamefont {A.}~\bibnamefont
  {Ourjoumtsev}}, \bibinfo {author} {\bibfnamefont {H.}~\bibnamefont {Jeong}},
  \bibinfo {author} {\bibfnamefont {R.}~\bibnamefont {Tualle-Brouri}},\ and\
  \bibinfo {author} {\bibfnamefont {P.}~\bibnamefont {Grangier}},\ }\bibfield
  {title} {\bibinfo {title} {Generation of optical `schr{\"o}dinger cats' from
  photon number states},\ }\href {https://doi.org/10.1038/nature06054}
  {\bibfield  {journal} {\bibinfo  {journal} {Nature}\ }\textbf {\bibinfo
  {volume} {448}},\ \bibinfo {pages} {784} (\bibinfo {year}
  {2007})}\BibitemShut {NoStop}%
\bibitem [{\citenamefont {Sychev}\ \emph {et~al.}(2017)\citenamefont {Sychev},
  \citenamefont {Ulanov}, \citenamefont {Pushkina}, \citenamefont {Richards},
  \citenamefont {Fedorov},\ and\ \citenamefont {Lvovsky}}]{Sychev2017}%
  \BibitemOpen
  \bibfield  {author} {\bibinfo {author} {\bibfnamefont {D.~V.}\ \bibnamefont
  {Sychev}}, \bibinfo {author} {\bibfnamefont {A.~E.}\ \bibnamefont {Ulanov}},
  \bibinfo {author} {\bibfnamefont {A.~A.}\ \bibnamefont {Pushkina}}, \bibinfo
  {author} {\bibfnamefont {M.~W.}\ \bibnamefont {Richards}}, \bibinfo {author}
  {\bibfnamefont {I.~A.}\ \bibnamefont {Fedorov}},\ and\ \bibinfo {author}
  {\bibfnamefont {A.~I.}\ \bibnamefont {Lvovsky}},\ }\bibfield  {title}
  {\bibinfo {title} {Enlargement of optical schr{\"o}dinger's cat states},\
  }\href {https://doi.org/10.1038/nphoton.2017.57} {\bibfield  {journal}
  {\bibinfo  {journal} {Nature Photonics}\ }\textbf {\bibinfo {volume} {11}},\
  \bibinfo {pages} {379} (\bibinfo {year} {2017})}\BibitemShut {NoStop}%
\bibitem [{\citenamefont {Takase}\ \emph {et~al.}(2021)\citenamefont {Takase},
  \citenamefont {Yoshikawa}, \citenamefont {Asavanant}, \citenamefont {Endo},\
  and\ \citenamefont {Furusawa}}]{PhysRevA.103.013710}%
  \BibitemOpen
  \bibfield  {author} {\bibinfo {author} {\bibfnamefont {K.}~\bibnamefont
  {Takase}}, \bibinfo {author} {\bibfnamefont {J.}~\bibnamefont {Yoshikawa}},
  \bibinfo {author} {\bibfnamefont {W.}~\bibnamefont {Asavanant}}, \bibinfo
  {author} {\bibfnamefont {M.}~\bibnamefont {Endo}},\ and\ \bibinfo {author}
  {\bibfnamefont {A.}~\bibnamefont {Furusawa}},\ }\bibfield  {title} {\bibinfo
  {title} {Generation of optical schr\"odinger cat states by generalized photon
  subtraction},\ }\href {https://doi.org/10.1103/PhysRevA.103.013710}
  {\bibfield  {journal} {\bibinfo  {journal} {Phys. Rev. A}\ }\textbf {\bibinfo
  {volume} {103}},\ \bibinfo {pages} {013710} (\bibinfo {year}
  {2021})}\BibitemShut {NoStop}%
\bibitem [{\citenamefont {Menicucci}(2011)}]{PhysRevA.83.062314}%
  \BibitemOpen
  \bibfield  {author} {\bibinfo {author} {\bibfnamefont {N.~C.}\ \bibnamefont
  {Menicucci}},\ }\bibfield  {title} {\bibinfo {title} {Temporal-mode
  continuous-variable cluster states using linear optics},\ }\href
  {https://doi.org/10.1103/PhysRevA.83.062314} {\bibfield  {journal} {\bibinfo
  {journal} {Phys. Rev. A}\ }\textbf {\bibinfo {volume} {83}},\ \bibinfo
  {pages} {062314} (\bibinfo {year} {2011})}\BibitemShut {NoStop}%
\bibitem [{\citenamefont {Lee}\ and\ \citenamefont
  {Nha}(2010)}]{PhysRevA.82.053812}%
  \BibitemOpen
  \bibfield  {author} {\bibinfo {author} {\bibfnamefont {S.-Y.}\ \bibnamefont
  {Lee}}\ and\ \bibinfo {author} {\bibfnamefont {H.}~\bibnamefont {Nha}},\
  }\bibfield  {title} {\bibinfo {title} {Quantum state engineering by a
  coherent superposition of photon subtraction and addition},\ }\href
  {https://doi.org/10.1103/PhysRevA.82.053812} {\bibfield  {journal} {\bibinfo
  {journal} {Phys. Rev. A}\ }\textbf {\bibinfo {volume} {82}},\ \bibinfo
  {pages} {053812} (\bibinfo {year} {2010})}\BibitemShut {NoStop}%
\bibitem [{\citenamefont {Vahlbruch}\ \emph {et~al.}(2016)\citenamefont
  {Vahlbruch}, \citenamefont {Mehmet}, \citenamefont {Danzmann},\ and\
  \citenamefont {Schnabel}}]{PhysRevLett.117.110801}%
  \BibitemOpen
  \bibfield  {author} {\bibinfo {author} {\bibfnamefont {H.}~\bibnamefont
  {Vahlbruch}}, \bibinfo {author} {\bibfnamefont {M.}~\bibnamefont {Mehmet}},
  \bibinfo {author} {\bibfnamefont {K.}~\bibnamefont {Danzmann}},\ and\
  \bibinfo {author} {\bibfnamefont {R.}~\bibnamefont {Schnabel}},\ }\bibfield
  {title} {\bibinfo {title} {Detection of 15 db squeezed states of light and
  their application for the absolute calibration of photoelectric quantum
  efficiency},\ }\href {https://doi.org/10.1103/PhysRevLett.117.110801}
  {\bibfield  {journal} {\bibinfo  {journal} {Phys. Rev. Lett.}\ }\textbf
  {\bibinfo {volume} {117}},\ \bibinfo {pages} {110801} (\bibinfo {year}
  {2016})}\BibitemShut {NoStop}%
\bibitem [{\citenamefont {Leghtas}\ \emph {et~al.}(2013)\citenamefont
  {Leghtas}, \citenamefont {Kirchmair}, \citenamefont {Vlastakis},
  \citenamefont {Schoelkopf}, \citenamefont {Devoret},\ and\ \citenamefont
  {Mirrahimi}}]{PhysRevLett.111.120501}%
  \BibitemOpen
  \bibfield  {author} {\bibinfo {author} {\bibfnamefont {Z.}~\bibnamefont
  {Leghtas}}, \bibinfo {author} {\bibfnamefont {G.}~\bibnamefont {Kirchmair}},
  \bibinfo {author} {\bibfnamefont {B.}~\bibnamefont {Vlastakis}}, \bibinfo
  {author} {\bibfnamefont {R.~J.}\ \bibnamefont {Schoelkopf}}, \bibinfo
  {author} {\bibfnamefont {M.~H.}\ \bibnamefont {Devoret}},\ and\ \bibinfo
  {author} {\bibfnamefont {M.}~\bibnamefont {Mirrahimi}},\ }\bibfield  {title}
  {\bibinfo {title} {Hardware-efficient autonomous quantum memory protection},\
  }\href {https://doi.org/10.1103/PhysRevLett.111.120501} {\bibfield  {journal}
  {\bibinfo  {journal} {Phys. Rev. Lett.}\ }\textbf {\bibinfo {volume} {111}},\
  \bibinfo {pages} {120501} (\bibinfo {year} {2013})}\BibitemShut {NoStop}%
\bibitem [{\citenamefont {Li}\ \emph {et~al.}(2017)\citenamefont {Li},
  \citenamefont {Zou}, \citenamefont {Albert}, \citenamefont {Muralidharan},
  \citenamefont {Girvin},\ and\ \citenamefont
  {Jiang}}]{PhysRevLett.119.030502}%
  \BibitemOpen
  \bibfield  {author} {\bibinfo {author} {\bibfnamefont {L.}~\bibnamefont
  {Li}}, \bibinfo {author} {\bibfnamefont {C.-L.}\ \bibnamefont {Zou}},
  \bibinfo {author} {\bibfnamefont {V.~V.}\ \bibnamefont {Albert}}, \bibinfo
  {author} {\bibfnamefont {S.}~\bibnamefont {Muralidharan}}, \bibinfo {author}
  {\bibfnamefont {S.~M.}\ \bibnamefont {Girvin}},\ and\ \bibinfo {author}
  {\bibfnamefont {L.}~\bibnamefont {Jiang}},\ }\bibfield  {title} {\bibinfo
  {title} {Cat codes with optimal decoherence suppression for a lossy bosonic
  channel},\ }\href {https://doi.org/10.1103/PhysRevLett.119.030502} {\bibfield
   {journal} {\bibinfo  {journal} {Phys. Rev. Lett.}\ }\textbf {\bibinfo
  {volume} {119}},\ \bibinfo {pages} {030502} (\bibinfo {year}
  {2017})}\BibitemShut {NoStop}%
\bibitem [{\citenamefont {Alexander}\ \emph {et~al.}(2014)\citenamefont
  {Alexander}, \citenamefont {Armstrong}, \citenamefont {Ukai},\ and\
  \citenamefont {Menicucci}}]{PhysRevA.90.062324}%
  \BibitemOpen
  \bibfield  {author} {\bibinfo {author} {\bibfnamefont {R.~N.}\ \bibnamefont
  {Alexander}}, \bibinfo {author} {\bibfnamefont {S.~C.}\ \bibnamefont
  {Armstrong}}, \bibinfo {author} {\bibfnamefont {R.}~\bibnamefont {Ukai}},\
  and\ \bibinfo {author} {\bibfnamefont {N.~C.}\ \bibnamefont {Menicucci}},\
  }\bibfield  {title} {\bibinfo {title} {Noise analysis of single-mode gaussian
  operations using continuous-variable cluster states},\ }\href
  {https://doi.org/10.1103/PhysRevA.90.062324} {\bibfield  {journal} {\bibinfo
  {journal} {Phys. Rev. A}\ }\textbf {\bibinfo {volume} {90}},\ \bibinfo
  {pages} {062324} (\bibinfo {year} {2014})}\BibitemShut {NoStop}%
\bibitem [{\citenamefont {Miyata}\ \emph {et~al.}(2016)\citenamefont {Miyata},
  \citenamefont {Ogawa}, \citenamefont {Marek}, \citenamefont {Filip},
  \citenamefont {Yonezawa}, \citenamefont {Yoshikawa},\ and\ \citenamefont
  {Furusawa}}]{PhysRevA.93.022301}%
  \BibitemOpen
  \bibfield  {author} {\bibinfo {author} {\bibfnamefont {K.}~\bibnamefont
  {Miyata}}, \bibinfo {author} {\bibfnamefont {H.}~\bibnamefont {Ogawa}},
  \bibinfo {author} {\bibfnamefont {P.}~\bibnamefont {Marek}}, \bibinfo
  {author} {\bibfnamefont {R.}~\bibnamefont {Filip}}, \bibinfo {author}
  {\bibfnamefont {H.}~\bibnamefont {Yonezawa}}, \bibinfo {author}
  {\bibfnamefont {J.}~\bibnamefont {Yoshikawa}},\ and\ \bibinfo {author}
  {\bibfnamefont {A.}~\bibnamefont {Furusawa}},\ }\bibfield  {title} {\bibinfo
  {title} {Implementation of a quantum cubic gate by an adaptive non-gaussian
  measurement},\ }\href {https://doi.org/10.1103/PhysRevA.93.022301} {\bibfield
   {journal} {\bibinfo  {journal} {Phys. Rev. A}\ }\textbf {\bibinfo {volume}
  {93}},\ \bibinfo {pages} {022301} (\bibinfo {year} {2016})}\BibitemShut
  {NoStop}%
\bibitem [{\citenamefont {Ghose}\ and\ \citenamefont
  {Sanders}(2007)}]{doi:10.1080/09500340601101575}%
  \BibitemOpen
  \bibfield  {author} {\bibinfo {author} {\bibfnamefont {S.}~\bibnamefont
  {Ghose}}\ and\ \bibinfo {author} {\bibfnamefont {B.~C.}\ \bibnamefont
  {Sanders}},\ }\bibfield  {title} {\bibinfo {title} {Non-gaussian ancilla
  states for continuous variable quantum computation via gaussian maps},\
  }\href {https://doi.org/10.1080/09500340601101575} {\bibfield  {journal}
  {\bibinfo  {journal} {Journal of Modern Optics}\ }\textbf {\bibinfo {volume}
  {54}},\ \bibinfo {pages} {855} (\bibinfo {year} {2007})},\ \Eprint
  {https://arxiv.org/abs/https://doi.org/10.1080/09500340601101575}
  {https://doi.org/10.1080/09500340601101575} \BibitemShut {NoStop}%
\bibitem [{\citenamefont {{Konno}}\ \emph {et~al.}(2020)\citenamefont
  {{Konno}}, \citenamefont {{Sakaguchi}}, \citenamefont {{Asavanant}},
  \citenamefont {{Ogawa}}, \citenamefont {{Kobayashi}}, \citenamefont
  {{Marek}}, \citenamefont {{Filip}}, \citenamefont {{Yoshikawa}},\ and\
  \citenamefont {{Furusawa}}}]{2020arXiv201114576K}%
  \BibitemOpen
  \bibfield  {author} {\bibinfo {author} {\bibfnamefont {S.}~\bibnamefont
  {{Konno}}}, \bibinfo {author} {\bibfnamefont {A.}~\bibnamefont
  {{Sakaguchi}}}, \bibinfo {author} {\bibfnamefont {W.}~\bibnamefont
  {{Asavanant}}}, \bibinfo {author} {\bibfnamefont {H.}~\bibnamefont
  {{Ogawa}}}, \bibinfo {author} {\bibfnamefont {M.}~\bibnamefont
  {{Kobayashi}}}, \bibinfo {author} {\bibfnamefont {P.}~\bibnamefont
  {{Marek}}}, \bibinfo {author} {\bibfnamefont {R.}~\bibnamefont {{Filip}}},
  \bibinfo {author} {\bibfnamefont {J.}~\bibnamefont {{Yoshikawa}}},\ and\
  \bibinfo {author} {\bibfnamefont {A.}~\bibnamefont {{Furusawa}}},\ }\bibfield
   {title} {\bibinfo {title} {{Nonlinear squeezing for measurement-based
  non-Gaussian operations in time domain}},\ }\href@noop {} {\bibfield
  {journal} {\bibinfo  {journal} {arXiv e-prints}\ ,\ \bibinfo {eid}
  {arXiv:2011.14576}} (\bibinfo {year} {2020})},\ \Eprint
  {https://arxiv.org/abs/2011.14576} {arXiv:2011.14576 [quant-ph]} \BibitemShut
  {NoStop}%
\bibitem [{\citenamefont {Arakawa}\ and\ \citenamefont
  {Holmes}(2020)}]{doi:10.1063/5.0010193}%
  \BibitemOpen
  \bibfield  {author} {\bibinfo {author} {\bibfnamefont {Y.}~\bibnamefont
  {Arakawa}}\ and\ \bibinfo {author} {\bibfnamefont {M.~J.}\ \bibnamefont
  {Holmes}},\ }\bibfield  {title} {\bibinfo {title} {Progress in quantum-dot
  single photon sources for quantum information technologies: A broad spectrum
  overview},\ }\href {https://doi.org/10.1063/5.0010193} {\bibfield  {journal}
  {\bibinfo  {journal} {Applied Physics Reviews}\ }\textbf {\bibinfo {volume}
  {7}},\ \bibinfo {pages} {021309} (\bibinfo {year} {2020})},\ \Eprint
  {https://arxiv.org/abs/https://doi.org/10.1063/5.0010193}
  {https://doi.org/10.1063/5.0010193} \BibitemShut {NoStop}%
\bibitem [{\citenamefont {Miwa}\ \emph {et~al.}(2014)\citenamefont {Miwa},
  \citenamefont {Yoshikawa}, \citenamefont {Iwata}, \citenamefont {Endo},
  \citenamefont {Marek}, \citenamefont {Filip}, \citenamefont {van Loock},\
  and\ \citenamefont {Furusawa}}]{PhysRevLett.113.013601}%
  \BibitemOpen
  \bibfield  {author} {\bibinfo {author} {\bibfnamefont {Y.}~\bibnamefont
  {Miwa}}, \bibinfo {author} {\bibfnamefont {J.}~\bibnamefont {Yoshikawa}},
  \bibinfo {author} {\bibfnamefont {N.}~\bibnamefont {Iwata}}, \bibinfo
  {author} {\bibfnamefont {M.}~\bibnamefont {Endo}}, \bibinfo {author}
  {\bibfnamefont {P.}~\bibnamefont {Marek}}, \bibinfo {author} {\bibfnamefont
  {R.}~\bibnamefont {Filip}}, \bibinfo {author} {\bibfnamefont
  {P.}~\bibnamefont {van Loock}},\ and\ \bibinfo {author} {\bibfnamefont
  {A.}~\bibnamefont {Furusawa}},\ }\bibfield  {title} {\bibinfo {title}
  {Exploring a new regime for processing optical qubits: Squeezing and
  unsqueezing single photons},\ }\href
  {https://doi.org/10.1103/PhysRevLett.113.013601} {\bibfield  {journal}
  {\bibinfo  {journal} {Phys. Rev. Lett.}\ }\textbf {\bibinfo {volume} {113}},\
  \bibinfo {pages} {013601} (\bibinfo {year} {2014})}\BibitemShut {NoStop}%
\bibitem [{\citenamefont {Takeda}\ \emph {et~al.}(2019)\citenamefont {Takeda},
  \citenamefont {Takase},\ and\ \citenamefont {Furusawa}}]{Takedaeaaw4530}%
  \BibitemOpen
  \bibfield  {author} {\bibinfo {author} {\bibfnamefont {S.}~\bibnamefont
  {Takeda}}, \bibinfo {author} {\bibfnamefont {K.}~\bibnamefont {Takase}},\
  and\ \bibinfo {author} {\bibfnamefont {A.}~\bibnamefont {Furusawa}},\
  }\bibfield  {title} {\bibinfo {title} {On-demand photonic entanglement
  synthesizer},\ }\bibfield  {journal} {\bibinfo  {journal} {Science Advances}\
  }\textbf {\bibinfo {volume} {5}},\ \href
  {https://doi.org/10.1126/sciadv.aaw4530} {10.1126/sciadv.aaw4530} (\bibinfo
  {year} {2019}),\ \Eprint
  {https://arxiv.org/abs/https://advances.sciencemag.org/content/5/5/eaaw4530.full.pdf}
  {https://advances.sciencemag.org/content/5/5/eaaw4530.full.pdf} \BibitemShut
  {NoStop}%
\bibitem [{\citenamefont {Larsen}\ \emph
  {et~al.}(2019{\natexlab{b}})\citenamefont {Larsen}, \citenamefont {Guo},
  \citenamefont {Breum}, \citenamefont {Neergaard-Nielsen},\ and\ \citenamefont
  {Andersen}}]{Larsen2019}%
  \BibitemOpen
  \bibfield  {author} {\bibinfo {author} {\bibfnamefont {M.~V.}\ \bibnamefont
  {Larsen}}, \bibinfo {author} {\bibfnamefont {X.}~\bibnamefont {Guo}},
  \bibinfo {author} {\bibfnamefont {C.~R.}\ \bibnamefont {Breum}}, \bibinfo
  {author} {\bibfnamefont {J.~S.}\ \bibnamefont {Neergaard-Nielsen}},\ and\
  \bibinfo {author} {\bibfnamefont {U.~L.}\ \bibnamefont {Andersen}},\
  }\bibfield  {title} {\bibinfo {title} {Fiber-coupled epr-state generation
  using a single temporally multiplexed squeezed light source},\ }\href
  {https://doi.org/10.1038/s41534-019-0170-y} {\bibfield  {journal} {\bibinfo
  {journal} {npj Quantum Information}\ }\textbf {\bibinfo {volume} {5}},\
  \bibinfo {pages} {46} (\bibinfo {year} {2019}{\natexlab{b}})}\BibitemShut
  {NoStop}%
\bibitem [{\citenamefont {Walschaers}\ \emph {et~al.}(2018)\citenamefont
  {Walschaers}, \citenamefont {Sarkar}, \citenamefont {Parigi},\ and\
  \citenamefont {Treps}}]{PhysRevLett.121.220501}%
  \BibitemOpen
  \bibfield  {author} {\bibinfo {author} {\bibfnamefont {M.}~\bibnamefont
  {Walschaers}}, \bibinfo {author} {\bibfnamefont {S.}~\bibnamefont {Sarkar}},
  \bibinfo {author} {\bibfnamefont {V.}~\bibnamefont {Parigi}},\ and\ \bibinfo
  {author} {\bibfnamefont {N.}~\bibnamefont {Treps}},\ }\bibfield  {title}
  {\bibinfo {title} {Tailoring non-gaussian continuous-variable graph states},\
  }\href {https://doi.org/10.1103/PhysRevLett.121.220501} {\bibfield  {journal}
  {\bibinfo  {journal} {Phys. Rev. Lett.}\ }\textbf {\bibinfo {volume} {121}},\
  \bibinfo {pages} {220501} (\bibinfo {year} {2018})}\BibitemShut {NoStop}%
\bibitem [{\citenamefont {Walschaers}\ and\ \citenamefont
  {Treps}(2020)}]{PhysRevLett.124.150501}%
  \BibitemOpen
  \bibfield  {author} {\bibinfo {author} {\bibfnamefont {M.}~\bibnamefont
  {Walschaers}}\ and\ \bibinfo {author} {\bibfnamefont {N.}~\bibnamefont
  {Treps}},\ }\bibfield  {title} {\bibinfo {title} {Remote generation of wigner
  negativity through einstein-podolsky-rosen steering},\ }\href
  {https://doi.org/10.1103/PhysRevLett.124.150501} {\bibfield  {journal}
  {\bibinfo  {journal} {Phys. Rev. Lett.}\ }\textbf {\bibinfo {volume} {124}},\
  \bibinfo {pages} {150501} (\bibinfo {year} {2020})}\BibitemShut {NoStop}%
\bibitem [{\citenamefont {Walschaers}\ \emph {et~al.}(2020)\citenamefont
  {Walschaers}, \citenamefont {Parigi},\ and\ \citenamefont
  {Treps}}]{PRXQuantum.1.020305}%
  \BibitemOpen
  \bibfield  {author} {\bibinfo {author} {\bibfnamefont {M.}~\bibnamefont
  {Walschaers}}, \bibinfo {author} {\bibfnamefont {V.}~\bibnamefont {Parigi}},\
  and\ \bibinfo {author} {\bibfnamefont {N.}~\bibnamefont {Treps}},\ }\bibfield
   {title} {\bibinfo {title} {Practical framework for conditional non-gaussian
  quantum state preparation},\ }\href
  {https://doi.org/10.1103/PRXQuantum.1.020305} {\bibfield  {journal} {\bibinfo
   {journal} {PRX Quantum}\ }\textbf {\bibinfo {volume} {1}},\ \bibinfo {pages}
  {020305} (\bibinfo {year} {2020})}\BibitemShut {NoStop}%
\bibitem [{\citenamefont {Ra}\ \emph {et~al.}(2020)\citenamefont {Ra},
  \citenamefont {Dufour}, \citenamefont {Walschaers}, \citenamefont {Jacquard},
  \citenamefont {Michel}, \citenamefont {Fabre},\ and\ \citenamefont
  {Treps}}]{Ra2020}%
  \BibitemOpen
  \bibfield  {author} {\bibinfo {author} {\bibfnamefont {Y.-S.}\ \bibnamefont
  {Ra}}, \bibinfo {author} {\bibfnamefont {A.}~\bibnamefont {Dufour}}, \bibinfo
  {author} {\bibfnamefont {M.}~\bibnamefont {Walschaers}}, \bibinfo {author}
  {\bibfnamefont {C.}~\bibnamefont {Jacquard}}, \bibinfo {author}
  {\bibfnamefont {T.}~\bibnamefont {Michel}}, \bibinfo {author} {\bibfnamefont
  {C.}~\bibnamefont {Fabre}},\ and\ \bibinfo {author} {\bibfnamefont
  {N.}~\bibnamefont {Treps}},\ }\bibfield  {title} {\bibinfo {title}
  {Non-gaussian quantum states of a multimode light field},\ }\href
  {https://doi.org/10.1038/s41567-019-0726-y} {\bibfield  {journal} {\bibinfo
  {journal} {Nature Physics}\ }\textbf {\bibinfo {volume} {16}},\ \bibinfo
  {pages} {144} (\bibinfo {year} {2020})}\BibitemShut {NoStop}%
\end{thebibliography}%

\end{document}